\documentclass[aps,prb,twocolumn,superscriptaddress,longbibliography]{revtex4-2}
\usepackage{amsmath,amssymb}
\usepackage[pdftex]{hyperref,graphicx}
\hypersetup{colorlinks = true, urlcolor = blue, linkcolor = blue, citecolor = blue}
\usepackage{xcolor}
\usepackage{bm}
\usepackage {verbatim}

\usepackage[etex=true,export]{adjustbox}
\usepackage{makecell}
\usepackage{multirow}
\usepackage{tabularx,multirow,array,diagbox}
\usepackage{adjustbox}
\usepackage{booktabs}

\newcommand{\beq}{\begin{equation}}
\newcommand{\eeq}{\end{equation}}
\newcommand{\beql}{\begin{equation*}}
\newcommand{\eeql}{\end{equation*}}
\newcommand{\beqn}{\begin{eqnarray}}
\newcommand{\eeqn}{\end{eqnarray}}

\begin{document}
\title{Symmetric Wannier states and tight-binding model for quantum spin Hall bands in $AB$-stacked MoTe$_2$/WSe$_2$ }
\author{Xun-Jiang Luo}
\affiliation{School of Physics and Technology, Wuhan University, Wuhan 430072, China}
\author{Minxuan Wang}
\affiliation{School of Physics and Technology, Wuhan University, Wuhan 430072, China}
\author{Fengcheng Wu}
\email{wufcheng@whu.edu.cn}
\affiliation{School of Physics and Technology, Wuhan University, Wuhan 430072, China}
\affiliation{Wuhan Institute of Quantum Technology, Wuhan 430206, China}

\begin{abstract}
Motivated by the observation of topological states in $AB$-stacked MoTe$_2$/WSe$_2$, we construct the symmetry-adapted Wannier states and tight-binding model for the quantum spin Hall bands in this system. Our construction is based on the symmetry analysis of Bloch states obtained from the continuum moir\'e Hamiltonian. For model parameters extracted from first-principles calculations, we find that the quantum spin Hall bands can be described by a tight-binding model defined on a triangular lattice with two Wannier states per site per valley. The two Wannier states in a given valley have the same Wannier center but
different angular momenta under threefold rotation.  The tight-binding model reproduces the energy spectrum and accurately describes the topological phase transition induced by the out-of-plane displacement field. Our study sheds light on the topological states in moir\'e  transition metal dichalcogenides bilayers, and provides a route to addressing the many-body physics in $AB$-stacked MoTe$_2$/WSe$_2$.
\end{abstract}
\maketitle

\section{introduction}
The discovery of correlated insulators and superconductors in magic angle twisted bilayer graphene \cite{Cao2018a,Cao2018} has demonstrated vast opportunities provided by moir\'e materials to design quantum phases of matter, including superconductors \cite{Xu2018,Po2018,Liu2018,Isobe2018,Wu2018a,Yankowitz2019,Wu2019a,Wu2019b,Lian2019,Chen2019a,Codecido2019,Shen2020,Saito2020,Stepanov2020}, correlated insulators \cite{Yuan2018,Wu2018,Koshino2018,Regan2020,Tang2020,Xie2020,Liu2020,Li2021a},  and nontrivial topological states \cite{Wu2017,Zou2018,Wu2019,Po2019,Song2019,Sharpe2019,Chen2020,Serlin2020,Chen2021}.  One promising direction is to study the interplay between many-body interactions and band topology, since moir\'e superlattices can often host topological flatbands with enhanced interaction effects. A prominent example of topological states is the quantum anomalous Hall insulator (QAHI), which has been realized in various graphene-based moir\'e systems \cite{Sharpe2019,Serlin2020,Chen2020,Polshyn2020,Nuckolls2020,Das2021,Stepanov2021,Polshyn2022}.

Moir\'e superlattices formed in bilayers of semiconducting transition metal dichalcogenides (TMD) can host moir\'e flatbands in a wider range of twist angles \cite{Wu2018}. Interaction-driven quantum phases \cite{Tang2020,Regan2020,Xu2020,Wang2020,Li2021a}  such as Mott insulators  and generalized Wigner crystals have been observed in moir\'e TMD bilayers. A theoretical work \cite{Wu2019} predicted that moir\'e bands in twisted TMD homobilayers can realize quantum spin Hall insulators (QSHI), which is possible because of the strong spin-orbit coupling in TMD. Although topological states have so far not been experimentally observed  in twisted TMD homobilayers, an experiment \cite{Li2021} on $AB$-stacked TMD heterobilayer MoTe$_2$/WSe$_2$ reported signatures of QSHI at filling factor $\nu = 2$ (two holes per moir\'e unit cell) and QAHI at $\nu=1$. Here both topological states were induced by an external out-of-plane displacement field. This experiment \cite{Li2021} is remarkable as it clearly demonstrates that distinct types of topological states can be realized within one system. 

The displacement field-induced topological moir\'e bands in AB-stacked MoTe$_2$/WSe$_2$ have been theoretically established by large-scale first-principles calculations \cite{Zhang2021}. The external displacement field induces topological band inversion between moir\'e bands derived, respectively, from MoTe$_2$ and WSe$_2$. The QSHI at $\nu = 2$ is consistent with the band structure calculations. A recent experiment further demonstrated that a small out-of-plane magnetic field drives the QSHI at $\nu=2$ into a Chern insulator \cite{Zhao2022}, which can also be understood within single-particle physics. On the other hand, the QAHI at $\nu=1$ is a manifestation of electron correlation effects in topological bands, since interaction-induced spontaneous time-reversal symmetry breaking is necessary for the QAHI. The exact nature of the QAHI at $\nu=1$ is under active study and different types of symmetry-breaking states are proposed \cite{Pan2022,Xie2022a,Devakul2022,Chang2022,Xie2022c,Xie2022,Dong2022}. An optical spectroscopy measurement suggested that the QAHI at $\nu=1$ in AB-stacked MoTe$_2$/WSe$_2$ is valley coherent rather than valley polarized \cite{Tao2022}, but the microscopic mechanism remains an open question. 

Tight-binding (TB) description of the topological moir\'e bands provides not only 
insights to the band structure, but also an important starting point to study the interaction physics. 
For $AB$ stacked MoTe$_2$/WSe$_2$, previous works \cite{Zhang2021,Rademaker2022Spin} proposed a tight-binding model without explicitly constructing the  Wannier states, where the proposed model is a generalization of the Kane-Mele model. Recently, several works \cite{Devakul2022,Dong2022} started from the interacting Kane-Mele model to study interaction-driven topological phases in AB-stacked MoTe$_2$/WSe$_2$. However, the tight-binding model description for the quantum spin Hall bands in this system remains an open question since the Wannier states have not been constructed in previous studies. 

In this paper, we construct the symmetry-adapted Wannier states and the effective TB model for the quantum spin Hall bands of AB-stacked MoTe$_2$/WSe$_2$. We perform a detailed symmetry analysis of the moir\'e Hamiltonian and the moir\'e bands. The symmetry eigenvalues of the Bloch states at high-symmetry momenta uniquely determine the center of the Wannier states \cite{bradlyn2017topological}. For model parameters extracted from the first-principles calculations \cite{Zhang2021}, we find that the Wannier states for AB-stacked MoTe$_2$/WSe$_2$ in the topological regime form an effective triangular lattice. We construct the Wannier states and the TB model defined on the triangular lattice. The constructed TB model is distinct from the generalized Kane-Mele model \cite{Zhang2021,Rademaker2022Spin}, but similar to the Bernevig-Hughes-Zhang model \cite{Bernevig2006}.
The TB model not only reproduces the energy spectrum of the moir\'e bands, but also accurately describes the topological phase transition induced by the displacement field.
Our TB model can be used for addressing the electron interaction effects in AB-stacked MoTe$_2$/WSe$_2$.

The rest of the paper is organized as follows. In Sec.~\ref{II}, we present the moir\'e Hamiltonian and the topological phase diagram characterized by valley Chern numbers. In Sec.~\ref{III}, we analyze the symmetries of the moir\'e Hamiltonian and Bloch states. In Sec.~\ref{IV}, we construct the symmetric Wannier states informed by the symmetry eigenvalues of Bloch states at high-symmetry momenta. In Sec.~\ref{V}, we build the TB model based on the obtained Wannier states. In Sec.~\ref{VI}, we conclude with a discussion and summary. Appendixes \ref{appendixI} – \ref{Appendix IV} complement the main text by providing additional technical details.

\section{moir\'e band structure}
\label{II}
\subsection{Moir\'e Hamiltonian}
\begin{figure}
\centering
\includegraphics[width=3in]{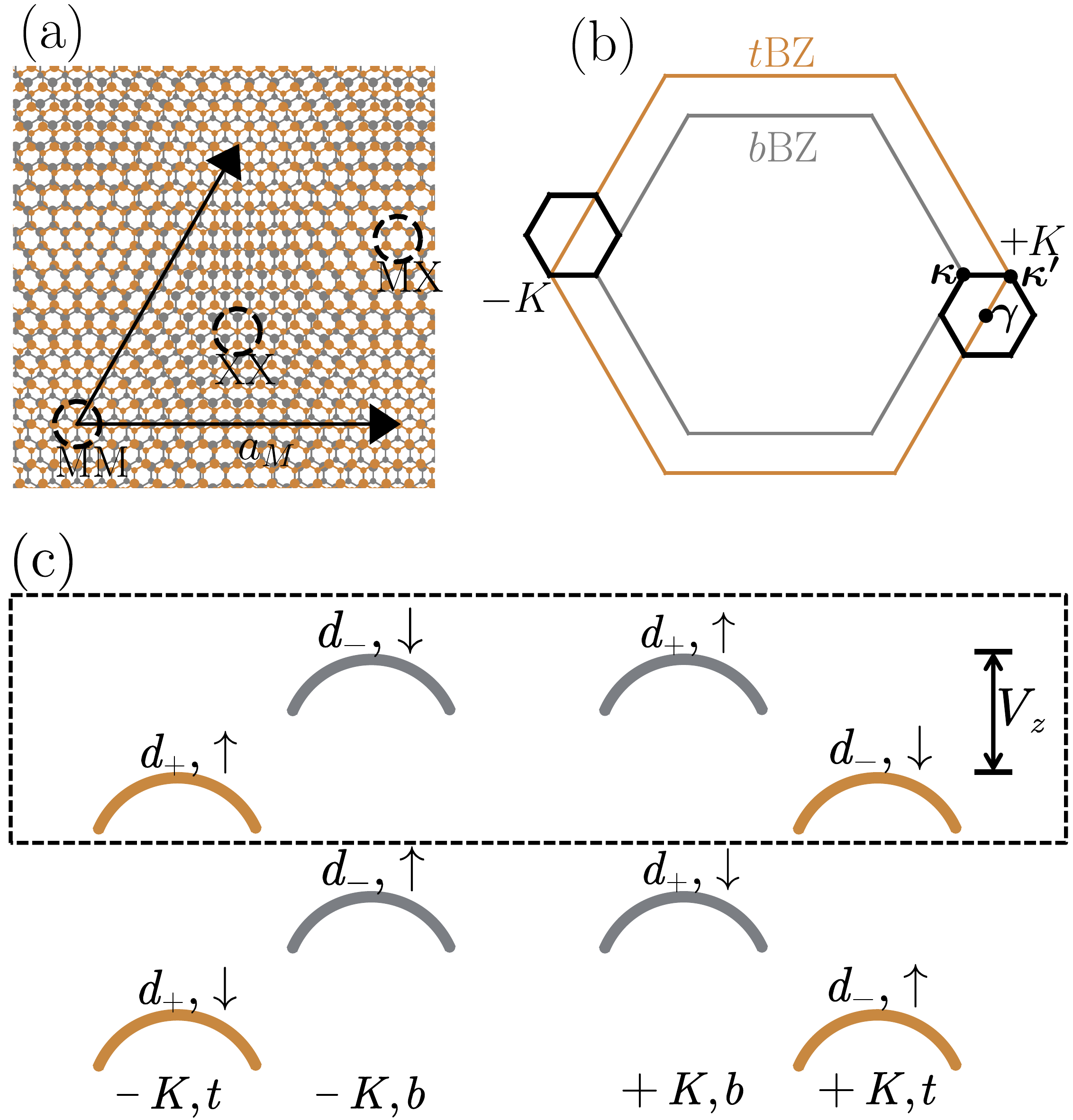}
\caption{(a) Moir\'e superlattices of AB-stacked MoTe$_2$/WSe$_2$ heterobilayer. (b) Schematic plot of the Brillouin zones. The gray and orange hexagons are the Brillouin zones of MoTe$_2$ and WSe$_2$, respectively. The left (right) black hexagon represents the moir\'e Brillouin zone in $-K$ ($+K$) valley.  (c) Schematic illustration of valence states in $\pm K$ valleys.  Only states in the dashed box are retained in the Hamiltonian $H_{\tau}$. }
\label{lattice}
\end{figure}

We study AB-stacked MoTe$_2$/WSe$_2$ with an exact 180$^\circ$ twist angle. The lattice constant mismatch generates a moir\'e superlattice with a period of $a_M=a_{{b}}a_{{t}}/|{a_{{b}}-a_{{t}}}|$, where $a_{{b}} = 3.575 \text{ \AA} $ and $a_{{t}} = 3.32 \text{ \AA} $ are the lattice constants of the bottom ($b$) MoTe$_2$ layer and the top ($t$) WSe$_2$ layer, respectively. The moir\'e superlattice, shown in Fig.~\ref{lattice}(a), has the $C_{3v}$ point group symmetry, which is generated by the threefold rotation around $z$ axis ($C_3$) and the mirror operation ($M_x$) that flips $x$ to $-x$. In the superlattice, there are three high-symmetry locations labeled by MM, XX, and MX. Here, MX refers to the location where the metal (M) atom of the bottom layer is vertically
aligned with the chalcogen atom (X) of the top layer, and likewise
for MM and XX locations. The momentum space structure is illustrated in Fig.~\ref{lattice}(b), which shows the Brillouin zones of each layer and the moir\'e superlattice.

The low-energy continuum Hamiltonian for AB-stacked MoTe$_2$/WSe$_2$ has been constructed in Ref.~\onlinecite{Zhang2021} informed by first-principles band structures and is given by
\begin{equation}\label{Ham}
H_{\tau}(\bm r)=\begin{pmatrix}
\mathcal{H}_{b,\tau}(\bm r)+\Delta_{{b}}(\bm{r}) & \Delta_{{T},\tau}(\bm{r})\\
\Delta_{{T},\tau}^\dag(\bm{r}) & \mathcal{H}_{t,\tau}(\bm r)+ \Delta_{t}(\bm{r})+V_{z}
\end{pmatrix},
\end{equation}
where $H_{\tau}$ is the valley-dependent moir\'e Hamiltonian for valence band states in $\tau K$ valley and $\tau=\pm 1$ is the valley index. 
Here $+K$ and $-K$ indicate corners of Brillouin zones associated with each monolayer and represent the valley degree of freedom. The valley index $\tau$ is a good quantum number in the low-energy Hamiltonian. 
As schematically demonstrated in Fig.~\ref{lattice}(c), the basis states of Hamiltonian $H_{\pm}$ are
\beqn
\begin{aligned}
\{|b,d_{+},\uparrow\rangle, |t,d_{-},\downarrow\rangle\} \,\,\,\,\,\,\text{for} \,\,\,\,\,\, H_{+},\\
\{ |b,d_{-},\downarrow\rangle, |t,d_{+},\uparrow\rangle\} \,\,\,\,\,\, \text{for} \,\,\,\,\,\, H_{-},
\end{aligned}
\eeqn
where $(b,t)$ are the layer indices, $|d_{\pm}\rangle=\frac{1}{\sqrt{2}}(|d_{x^2-y^2}\rangle\pm i|d_{xy}\rangle)$ represent the predominant atomic $d-$orbitals of the metal atoms, and $(\uparrow,\downarrow)$ are, respectively, for spin up and down. In a given valley, the basis states have layer-contrast orbital and spin characters, which is a result of the 180$^\circ$ rotation between the two layers.

$\mathcal{H}_{b,\tau}$ and $ \mathcal{H}_{t,\tau}$ in Eq.~\eqref{Ham} represent, respectively, the kinetic energy for the bottom and top layers,
\begin{equation}
\begin{aligned}
\mathcal{H}_{b,\tau} & =-\frac{\hbar^2\left(\hat{\bm{k}}-\tau \bm\kappa \right)^2}{2m_{b}},\\
\mathcal{H}_{t,\tau} &=-\frac{{\hbar^2\left(\hat{\bm{k}}-\tau \bm \kappa^{\prime} \right)^2}}{{2m_{t}}},
\end{aligned}
\end{equation}
where
$\hat{\bm k}=-i\partial_{\bm r}$ is the momentum operator, $\bm\kappa=(4\pi/3a_M)(-1/2,\sqrt{3}/2)$, $\bm\kappa^{\prime}=(4\pi/3a_M)(1/2,\sqrt{3}/2)$, and $(m_b,m_t)=(0.65 m_e, 0.35 m_e)$ are the effective masses for the two layers ($m_e$ is the rest electron mass). The momenta $\bm\kappa$ and $\bm\kappa^{\prime}$ are located at the corners of the moir\'e Brillouin zone and account for the momentum shift of the band extrema associated with the two layers (Fig.~\ref{lattice}(b)).

\begin{figure}
\centering
\includegraphics[width=3.5in]{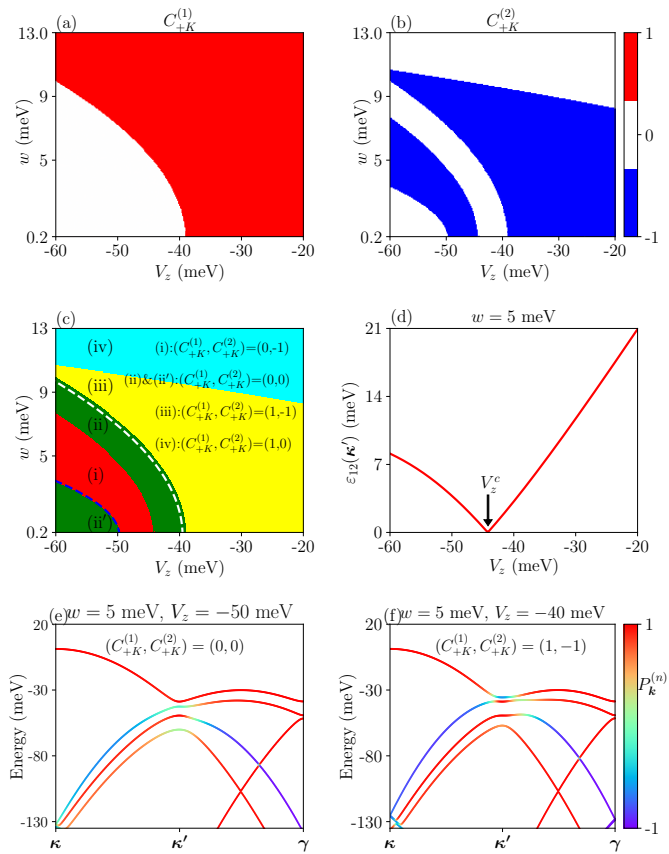}
\caption{(a), (b) The Chern numbers of the first and second moir\'e band at $+K$ valley in the parameters space $(V_z,w)$.   (c) The phase diagram characterized by the Chern numbers $(C_{+K}^{(1)},C_{+K}^{(2)})$.
The white dashed line obtained from Eq.~\eqref{Vz} represents an analytical approximation for the phase boundary between (ii) and (iii). Similarly, the blue dashed line obtained from Eq.~\eqref{eqv} closely follows the numerical phase boundary between (ii$^{\prime}$) and (i). (d) The energy gap $\varepsilon_{12}(\bm \kappa^{\prime})$ between the first and the second bands at $\bm \kappa^{\prime}$ point as a function of $V_z$ at $w=5$ meV.  (e), (f) The moir\'e bands in phases (ii) and (iii) with different values of $V_z$ at $w=5$ meV. }
\label{Chern}
\end{figure}

$\Delta_{b,t}$ and $\Delta_{{T},\tau}$ in Eq.~\eqref{Ham} are, respectively, the intralayer potential and interlayer tunneling, which are parametrized as follows,
\beqn
\begin{aligned}
 &\Delta_{{b,t}}(\bm{r})=2V_{{b,t}}\sum_{j=1,3,5} \cos(\bm{g}_j \cdot \bm{r}+\phi_{{b,t}}),\\
& \Delta_{{T},\tau}(\bm{r})=\tau w \left(1+\omega^{\tau} e^{i\tau\bm{g}_2\cdot\bm{r}}+\omega^{2\tau} e^{i\tau\bm{g}_3\cdot\bm{r}} \right),
\end{aligned}
\eeqn
where $(V_{b,t},\phi_{{b,t}},w)$ are model parameters, and $\bm{g}_j=4\pi/(\sqrt{3}a_M)\{-\sin [\pi(j-1)/3],\cos[\pi(j-1)/3]\}$ are the moir\'e reciprocal
lattice vectors in the first shell. The form of $\Delta_{b,t}$ and $\Delta_{{T},\tau}$ is constrained by symmetry. In particular, the phase factor $\omega$ is fixed to be $e^{i 2\pi/3}$ by the threefold rotation symmetry $C_3$. Therefore, the tunneling term $\Delta_{{T},\tau}$ has a finite value at the XX location, but vanishes at the MM and MX locations.

$V_z$ in Eq.~\eqref{Ham} is the band offset between different layers and can be tuned by an applied vertical displacement field. At zero displacement field, the intrinsic band offset $V_z$ is around $-110$ meV \cite{Zhang2021}. The other model parameters have been determined in Ref.~\onlinecite{Zhang2021} from fitting to the first-principles band structures, and take the following values, $V_b = 4.1$ meV, $\phi_b=14^{\circ}$, $V_t =0$, and $w= 1.3$ meV. Here, $V_t$ is set to be 0, because the low-energy physics only involves the valence band maximum of WSe$_2$, and the potential $\Delta_t(\bm r)$ can be neglected.  
 We note that the first-principles calculation might not be accurate enough to precisely determine $w$ that is on the scale of 1 meV.
Experimentally, $w$ could be modified by pressure \cite{Yankowitz2019}. Therefore, we take $w$ and $V_z$ as adjustable theoretical parameters to study the topological phase diagram, but keep the values of other parameters fixed.

The moir\'e Hamiltonians $H_{+}$ and $H_{-}$ are related by the time-reversal symmetry $\mathcal{T}=i\tau_y\sigma_z\mathcal{K}$, where $\sigma_z$ and $\tau_y$ are Pauli matrices in the layer and valley spaces, and $\mathcal{K}$ is the complex conjugation operator. We present a detailed discussion of the $\mathcal{T}$ symmetry in Appendix~\ref{appendixI}. In the following, we mainly focus on the physics of $H_{+}$ in $+K$ valley, unless otherwise stated.

\begin{figure*}
\centering
\includegraphics[width=0.98\textwidth]{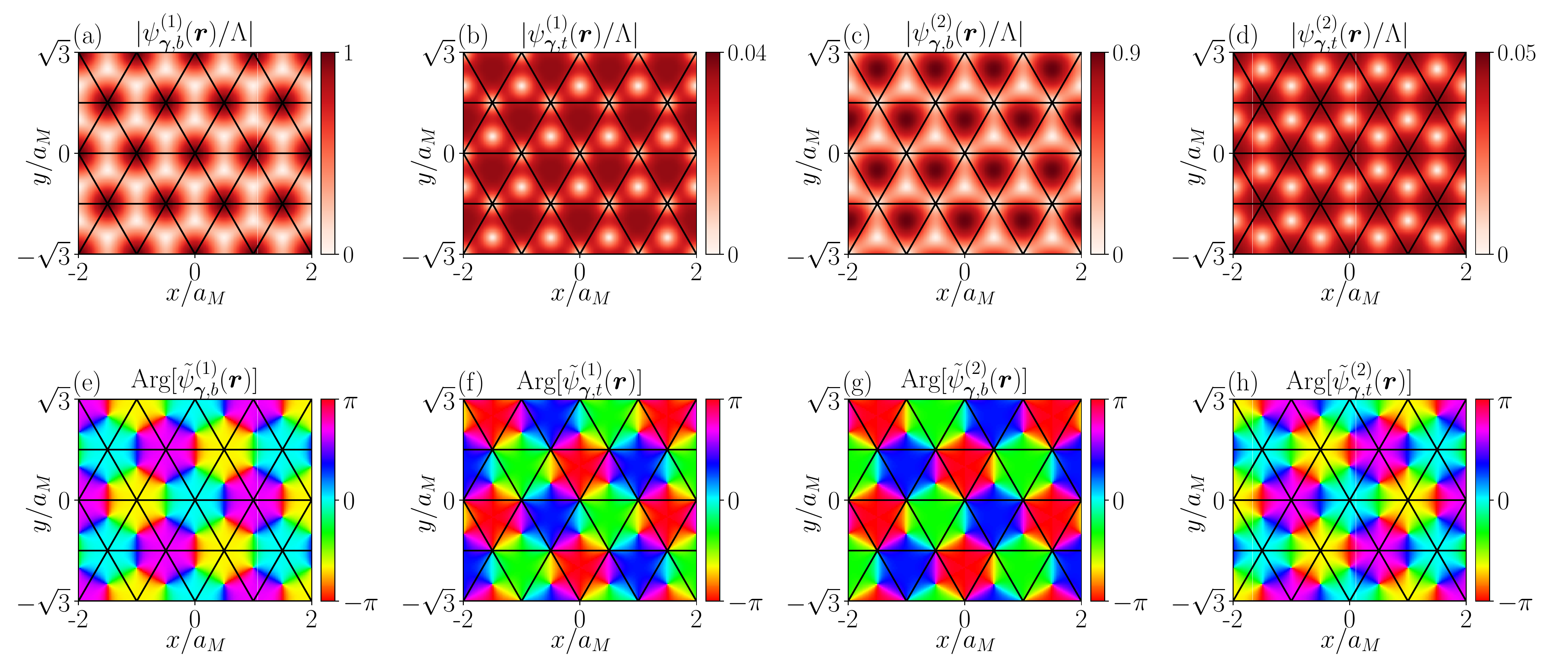}
\caption{The amplitude and phase of states ${\psi}_{\bm \gamma}^{(n)}(\bm r)=[{\psi}_{\bm \gamma,b}^{(n)}(\bm r),{\psi}_{\bm \gamma,t}^{(n)}(\bm r)]^{T}$ at $\bm \gamma$ point. (a)-(d) The amplitude of ${\psi}_{\bm \gamma,l}^{(n)}(\bm r)/\Lambda$,  where $l=b,t$ is the layer index, and $\Lambda=\psi_{\bm \gamma,b}^{(1)}(\bm r=0)$ is a normalization factor. (e)-(h) The phase of
$\tilde{\psi}_{\bm \gamma,l}^{(n)}(\bm r)$, where $\tilde{\psi}_{\bm \gamma,b}^{(n)}(\bm r)=e^{-i\bm\kappa\bm\cdot \bm r}\psi_{\bm\gamma,b}^{(n)}(\bm r)$ and $\tilde{\psi}_{\bm \gamma,t}^{(n)}(\bm r)=e^{-i\bm\kappa^{\prime}\bm\cdot \bm r}\psi_{\bm\gamma,t}^{(n)}(\bm r)$.  The black lines mark the effective
triangular lattice formed by the MM points. Parameter values are the same as those used for Fig.~\ref{Chern}(f).  }
\label{Bloch}
\end{figure*}

\subsection{Topological phase diagram}

The topology of the moir\'e bands can be tuned by the band offset $V_z$. In
the intrinsic case without external displacement field $(V_z \sim -110 \text{meV})$, the topmost moir\'e valence bands are mainly derived from
the MoTe$_2$ layer and topologically trivial. When $|V_z|$ is reduced by an applied displacement field, there can be band inversion between bands derived from different layers, which can drive topological phase transitions \cite{Zhang2021,Pan2022}.

To characterize the band topology, we calculate $C_{+K}^{(1)}$ and $C_{+K}^{(2)}$ in the parameter space of $(V_z,w)$, as shown in Figs.~\ref{Chern}(a) and \ref{Chern}(b), respectively. Here $C_{+K}^{(n)}$ is the Chern number of the $n$th moir\'e valance band at $+K$ valley. Based on the Chern numbers, the parameter space $(V_z,w)$ in Fig.~\ref{Chern}(c) can be classified into five regions; $(C_{+K}^{(1)},C_{+K}^{(2)})$ take values of $(0,-1)$ in phase (i), $(0,0)$ in phases (ii) and (ii$^\prime$), $(1,-1)$ in phase (iii), and $(1,0)$ in phase (iv), respectively. Here phases (ii) and (ii$^\prime$) have identical Chern numbers for the first two bands, but we use the two different labels to emphasize that they are separated in the parameter space by phase (i). Phase (ii$^\prime$) appears in the lower-left corner of the parameter space with weak $w$ and sufficiently negative $V_z$, where both of the first two bands are mainly derived from the bottom layer and topologically trivial. 

In this work, we focus particularly on phase (iii), since the first moir\'e valence bands in this phase realize the quantum spin Hall state when both valleys are considered. Because the valley index is a good quantum number in our low-energy continuum model and the $\pm K$ valleys are related by time-reversal symmetry, 
the $Z_2$ topological invariant for the quantum spin Hall state can be defined as $Z_2=(C_{+K}^{(1)}-C_{-K}^{(1)})/2$ mod 2. The valley Chern numbers are related by time-reversal symmetry as $C_{-K}^{(1)}=-C_{+K}^{(1)}$. Therefore, the $Z_2$ invariant is  1 (nontrivial)  for the first moir\'e bands in phase (iii). Note that phase (iv) also generates the quantum spin Hall state in the first moir\'e valence bands. However, its existence requires a value of $w$ that is possibly too large for AB-stacked MoTe$_2$/WSe$_2$. We keep phase (iv) in the phase diagram for completeness but do not study it further in this work.

Phase (ii) is separated from phase (iii) by a topological phase transition tuned by $V_z$. At the critical point $V_z=V_z^{c}$ for the transition, the energy gap $\varepsilon_{12}(\bm \kappa^{\prime})$ between the first and the second moir\'e bands closes at $\bm\kappa^{\prime}$ point in $+K$ valley, as shown in Fig.~\ref{Chern}(d).
This phase transition is further revealed by the moir\'e band structures in Figs.~\ref{Chern}(e) and \ref{Chern}(f) for phases (ii) and (iii), respectively. The color of the bands encodes the layer polarization $P_{\bm k}^{(n)}$, which is defined by
\beqn
P_{\bm k}^{(n)}=\langle \psi_{\bm k}^{(n)}|\sigma_z|\psi_{\bm k}^{(n)}\rangle.
\label{lp}
\eeqn
Here $\psi_{\bm k}^{(n)}$ is the Bloch state for the $n$th band at momentum $\bm k$ and is obtained by diagonalizing the moir\'e Hamiltonian $H_{+}(\bm r)$ in plane wave basis. In the layer pseudospin space, $\psi_{\bm k}^{(n)}$ is a two-component spinor $[\psi_{\bm k,{b}}^{(n)}, \psi_{\bm k,{t}}^{(n)}]^{T}$. The layer polarization clearly reveals the topological phase transition. When $V_z<V_z^{c}$,  $P_{\bm k}^{(1)}$ approaches 1 at every $\bm k$, indicating that the first band is mainly derived from MoTe$_2$ layer. After the topological phase transition ($V_z>V_z^{c}$), both $P_{\bm k}^{(1)}$ and $P_{\bm k}^{(2)}$ change sign for $\bm k$ around $\bm \kappa'$ point. Therefore, the band inversion at $\bm \kappa'$ point, which drives the topological phase transition, is characterized by the layer inversion.

\begin{figure*}
\centering
\includegraphics[width=0.98\textwidth]{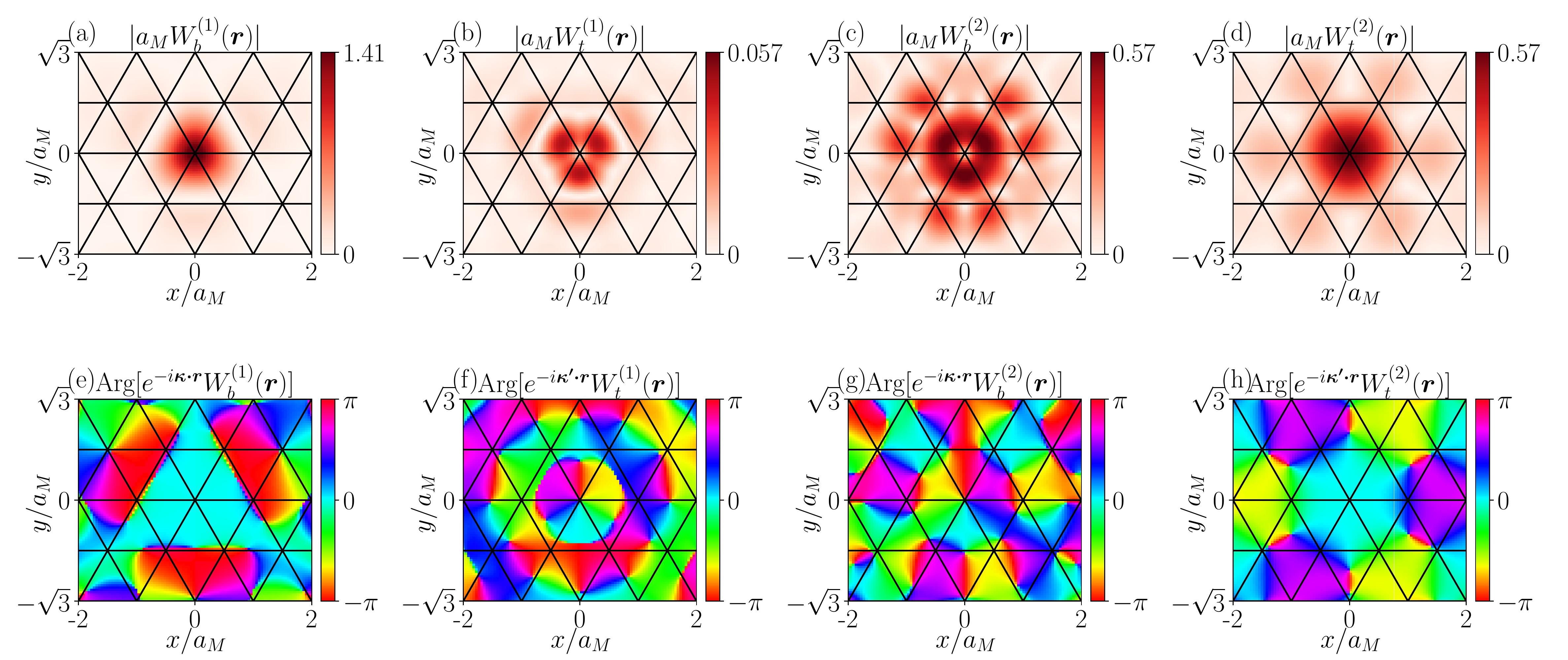}
\caption{The amplitude and phase of Wannier states ${W}^{(n)}(\bm r)=[{W}_{b}^{(n)}(\bm r),{W}_{t}^{(n)}(\bm r)]^{T}$. (a)-(d) The amplitude of ${W}_{l}^{(n)}(\bm r)$, where $l=b,t$ is the layer index.  (e)-(h) The phase of $e^{-i\bm \kappa\cdot\bm r}{W}_{b}^{(n)}(\bm r)$ and $e^{-i\bm \kappa^{\prime}\cdot\bm r}{W}_{t}^{(n)}(\bm r)$. We take the gauge such that ${W}_{b}^{(1)}(\bm r)$ and ${W}_{t}^{(2)}(\bm r)$ are real and positive at $\bm r=0$. The black lines mark the effective
triangular lattice. The parameter values are the same as those used for Fig.~\ref{Chern}(f).  }
\label{Wannier}
\end{figure*}

\subsection{Analytical phase boundaries}
To gain a deeper insight into the topological phase diagram, we construct an analytical theory for the phase boundary between phases (ii) and (iii). The approximate analytical theory is derived by truncating the moir\'e Hamiltonian at $\bm\kappa^{\prime}$ in the plane-wave basis to the first shell. In this approximation, we keep the following four low-energy plane-wave states,  $\left\{\left|\boldsymbol{\kappa}^{\prime}, b\right\rangle,\left|\boldsymbol{\kappa}^{\prime}+\boldsymbol{g}_2, b\right\rangle,\left|\boldsymbol{\kappa}^{\prime}+\boldsymbol{g}_3, b\right\rangle,\left|\boldsymbol{\kappa}^{\prime}, t\right\rangle\right\}$, where $b$ and $t$ refer to the layer degree of freedom. In the basis of these four states, the moir\'e Hamiltonian is
\beqn
H_{\boldsymbol{\kappa}^{\prime}, \tau=+} \approx\left(\begin{array}{cccc}
-E_\kappa & V_b e^{i \phi_b} & V_b e^{-i \phi_b} & w \\
V_b e^{-i \phi_b} & -E_\kappa & V_b e^{i \phi_b} & w e^{\frac{i 2 \pi}{3}} \\
V_b e^{i \phi_b} & V_b e^{-i \phi_b} & -E_\kappa & w e^{-\frac{i 2 \pi}{3}} \\
w & w e^{-\frac{i 2 \pi}{3}} & w e^{\frac{i 2 \pi}{3}} & V_z
\end{array}\right),
\label{Hkppap}
\eeqn
where $E_\kappa=\frac{\hbar^2\left|\boldsymbol{\kappa}^{\prime}-\boldsymbol{\kappa}\right|^2}{2 m_b}=\frac{\hbar^2|\boldsymbol{\kappa}|^2}{2 m_b}$. $  H_{\boldsymbol{\kappa}^{\prime}, \tau=+}$ can be block diagonalized by applying the following unitary transformation,
\beqn
&&\Lambda=\begin{pmatrix}
\frac{1}{\sqrt{3}} & \frac{1}{\sqrt{3}} & \frac{1}{\sqrt{3}} & 0 \\
\frac{1}{\sqrt{3}} & \frac{1}{\sqrt{3}} e^{-\frac{i 2 \pi}{3}} & \frac{1}{\sqrt{3}} e^{+\frac{i 2 \pi}{3}} & 0 \\
\frac{1}{\sqrt{3}} & \frac{1}{\sqrt{3}} e^{+\frac{i 2 \pi}{3}} & \frac{1}{\sqrt{3}} e^{-\frac{i 2 \pi}{3}} & 0 \\
0 & 0 & 0 & 1
\end{pmatrix}, \nonumber\\
&&\Lambda^{+} H_{\boldsymbol{\kappa}^{\prime}, \tau=+} \Lambda=\begin{pmatrix}
e_0 & 0 & 0 & 0 \\
0 & e_{-1} & 0 & 0 \\
0 & 0 & e_1 & \sqrt{3} w \\
0 & 0 & \sqrt{3} w & V_z
\end{pmatrix},
\eeqn
where $e_n=-E_\kappa+2 V_b \cos \left(\phi_b+\frac{2 \pi n}{3}\right)$ with $n=0$ and $\pm 1 $. The eigenvalues of $H_{\boldsymbol{\kappa}^{\prime}, \tau=+}$ are $e_0, e_{-1}, \frac{e_1+V_z}{2} \pm \sqrt{3 w^2+\left(\frac{e_1-V_z}{2}\right)^2}$. The gap $\varepsilon_{12}\left(\boldsymbol{\kappa}^{\prime}\right)$ closes when
\beqn
e_0=\frac{e_1+V_z}{2}+\sqrt{3 w^2+\left(\frac{e_1-V_z}{2}\right)^2},
\eeqn
which leads to an analytical expression for $V_z^c$,
\beqn
V_z^c=e_0-\frac{3 w^2}{e_0-e_1}.
\label{Vz}
\eeqn
Equation \eqref{Vz} agrees well with the numerical phase boundary between phases (ii) and (iii), as shown by the white dashed line in Fig.~\ref{Chern}(c).

The Hamiltonian $H_{\boldsymbol{\kappa}^{\prime}, \tau=+}$ in Eq.~\eqref{Hkppap} also captures the transition from phase $\left(\mathrm{ii}^{\prime}\right)$ to (i), which is signaled by the closing of the energy gap $\varepsilon_{23}\left(\boldsymbol{\kappa}^{\prime}\right)$ between the second and third bands  at the  $\boldsymbol{\kappa}^{\prime}$ point. Using the eigenvalues of $H_{\boldsymbol{\kappa}^{\prime}, \tau=+}$, we find that $\varepsilon_{23}\left(\boldsymbol{\kappa}^{\prime}\right)$ closes when
\beqn
e_{-1}=\frac{e_1+V_z}{2}+\sqrt{3 w^2+\left(\frac{e_1-V_z}{2}\right)^2},
\eeqn
which leads to another critical $V_z$,
\beqn
V_z^{c'}=e_{-1}-\frac{3 w^2}{e_{-1}-e_1}.
\label{eqv}
\eeqn
Equation (\ref{eqv}), which is represented by the blue dashed line in Fig.~\ref{Chern}(c), agrees excellently with the numerical phase boundary between phases $(\text{ii}^{\prime})$ and (i).

The transition between phases (i) and (ii) is accompanied by the closing of the energy gap $\varepsilon_{23}(\boldsymbol{\kappa})$ between the second and third bands at the $\boldsymbol{\kappa}$ point.  An approximate expression for $\varepsilon_{23}(\boldsymbol{\kappa})$ would require truncating the moir\'e Hamiltonian at $\boldsymbol{\kappa}$ in the plane-wave basis to the second shell; keeping more states would complicate the analysis, and therefore, we do not pursue to derive an analytical theory for the phase boundary between phases (i) and (ii).

\section{Symmetry}
\label{III}

We study the symmetry properties of the Hamiltonian and the Bloch states. 
At high-symmetry points in the Brillouin zone, the Bloch states are classified by the symmetry group of the system. The symmetry representations of the bands at the high-symmetry momenta play an essential role in determining whether and how the bands can be decomposed into symmetric Wannier orbitals \cite{bradlyn2017topological}. For example, in twisted bilayer graphene, the symmetry representations of the two low-energy bands near the charge neutrality point do not match with those of any atomic insulator, which leads to Wannier obstructions \cite{Po2018, Song2019}

For AB-stacked MoTe$_2$/WSe$_2$, symmetries include the threefold rotation $C_3$, the mirror operation $M_x$, and the time-reversal symmetry $\mathcal{T}$. The $C_3$ operation acts within one valley, while the $M_x$ and $\mathcal{T}$ operations change the valley index. However, the combined operation $M_x\mathcal{T}$ does not change the valley index.  In the following, we analyze the $C_3$ and $M_x\mathcal{T}$ symmetries of $H_{+}(\bm r)$ separately.

To study the ${C}_3$ symmetry, we first apply a unitary transformation to ${H}_{+}( \bm r)$,
\beqn
\begin{aligned}
& \tilde{H}_{+}(\bm r)\equiv U(\bm r)H_{+}(\bm r)U^{-1}(\bm r),\\
&U(\bm r)=\begin{pmatrix} e^{-i \bm \kappa\bm \cdot\bm r}& 0\\
0& e^{-i \bm \kappa^{\prime}\bm \cdot\bm r}\end{pmatrix},\\
&\tilde{H}_{+}(\bm r)=\begin{pmatrix}
-\frac{\hbar^2 \hat{\bm k}^2}{2m_b}+\Delta_b(\bm r)&\tilde{\Delta}_{{T}}(\bm r)\\
\tilde{\Delta}_{{T}}^{\dagger}(\bm r)&-\frac{\hbar^2 \hat{\bm k}^2}{2m_t}+\Delta_t(\bm r)+V_z
\end{pmatrix},
\label{C31}
\end{aligned}
\eeqn
where $\tilde{\Delta}_{{T}}(\bm r)=w(e^{i\bm {q_1\cdot r}}+e^{i2\pi/3}e^{i\bm q_2\bm \cdot\bm r}+e^{i4\pi/3}e^{i\bm q_3\bm \cdot\bm r})$. Here $\bm q_1=\bm \kappa^{\prime}-\bm \kappa,\bm q_2=\hat{R}_{3}\bm q_1$,  $\bm q_3=\hat{R}_{3}\bm q_2$, and $\hat{R}_{3}$ is the anticlockwise rotation by $2\pi/3$.
The new Hamiltonian $\tilde{H}_{+}(\bm r)$ has a transparent threefold rotation symmetry represented by $\tilde{C}_3$,
\beqn
\begin{aligned}
&\tilde{C}_3\tilde{H}_{+}(\bm r)\tilde{C}_3^{-1}\equiv D_{\tilde{C}_3}\tilde{H}_{+}(\hat{R}_{3}\bm r)D_{\tilde{C}_3}^{-1}, \\
&D_{\tilde{C}_3}=\begin{pmatrix} 1& 0\\
0& e^{i\frac{2\pi}{3}}\end{pmatrix},
\label{C32}
\end{aligned}
\eeqn
where $\tilde{C}_3$ not only rotates $\bm r$ to $\hat{R}_{3}\bm r$, but also includes a unitary transformation $D_{\tilde{C}_3}$. Here $D_{\tilde{C}_3}$ is determined (up to an arbitrary phase) by requiring that $\tilde{C}_3\tilde{H}_{+}(\bm r)\tilde{C}_3^{-1}=\tilde{H}_{+}(\bm r)$.

The ${C}_3$ symmetry of the Hamiltonian ${H}_{+}(\bm r)$ is, therefore, represented by $C_3=U^{-1}(\bm r) \tilde{C}_3 U(\bm r)$, and  acts on the Bloch state $\psi_{\bm k}^{(n)}$ in the following way,
\beqn
C_3\psi_{\bm k}^{(n)}(\bm r)&=&U^{-1}(\bm r)\tilde{C_3}\tilde{\psi}_{\bm k}^{(n)}(\bm r)\nonumber\\
&=&U^{-1}(\bm r)D_{\tilde{C}_3}\tilde{\psi}_{\bm k}^{(n)}(\hat{R}_3\bm r),
\label{C34}
\eeqn
where $\tilde{\psi}_{\bm k}^{(n)}(\bm r)=U(\bm r){\psi}_{\bm k}^{(n)}(\bm r)$. In the layer pseudospin space, $\tilde{\psi}_{\bm k}^{(n)}(\bm r)=[\tilde{\psi}_{\bm k,{b}}^{(n)}(\bm r),\tilde{\psi}_{\bm k,{t}}^{(n)}(\bm r)]^{T}$, where the two components are, respectively, given by
\beqn
\tilde{\psi}_{\bm k,{b}}^{(n)}(\bm r)=e^{-i\bm \kappa \bm \cdot \bm r}{\psi}_{\bm k,{b}}^{(n)}(\bm r),\quad\tilde{\psi}_{\bm k,{t}}^{(n)}(\bm r)=e^{-i\bm \kappa^{\prime} \bm \cdot \bm r}{\psi}_{\bm k,{t}}^{(n)}(\bm r).\nonumber\\
\eeqn

In the moir\'e Brillouin zone, there are three high-symmetry momenta $\bm\kappa$, $\bm\kappa^{\prime}$, and $\bm \gamma=(0,0)$, which are invariant under the threefold rotation. For $\bm k$ at one of these three momenta, ${\psi}_{\bm k}^{(n)}$ is the eigenstate of the symmetry operator ${C}_3$, 
\beqn
{C}_3{\psi}_{\bm k}^{(n)}(\bm r)=e^{i2\pi L_{\bm k}^{(n)}/3}{\psi}_{\bm k}^{(n)}(\bm r),
\label{C36}
\eeqn
where  $L_{\bm k}^{(n)}$ is the angular momentum of ${\psi}_{\bm k}^{(n)}$ under threefold rotation and is defined modulo 3. By combining Eqs.~\eqref{C34} and \eqref{C36}, we have
\beqn
D_{\tilde{C}_3}\tilde{\psi}_{\bm k}^{(n)}(\hat{R}_3\bm r)=e^{i2\pi L_{\bm k}^{(n)}/3}\tilde{\psi}_{\bm k}^{(n)}(\bm r).
\label{C37}
\eeqn

We now take the first moir\'e band in Fig.~\ref{Chern}(f) at $\bm \gamma$ point as an example to demonstrate the derivation of $L_{\bm\gamma}^{(1)}$. Figure~\ref{Bloch} plots the amplitude and phase for each layer component of $\tilde{\psi}_{\bm \gamma}^{(n)}(\bm r)$. 
Extracting the phase information from Figs.~\ref{Bloch}(e) and \ref{Bloch}(f), we find that
\beqn
\tilde{\psi}_{\bm\gamma,b}^{(1)}(\hat{R}_3\bm r)=\tilde{\psi}_{\bm\gamma,b}^{(1)}(\bm r),\quad \tilde{\psi}_{\bm\gamma,t}^{(1)}(\hat{R}_3\bm r)=e^{-i2\pi/3}\tilde{\psi}_{\bm\gamma,t}^{(1)}(\bm r).\nonumber\\
\eeqn
Thus, following Eq.~\eqref{C37}, we have
\beqn
\begin{aligned}
D_{\tilde{C}_3}\tilde{\psi}_{\bm \gamma}^{(1)}(\hat{R}_3\bm r)&=D_{\tilde{C}_3}\left(\begin{array}{c}\tilde{\psi}_{\bm\gamma,b}^{(1)}(\bm r)\\ e^{-i2\pi/3} \tilde{\psi}_{\bm\gamma,t}^{(1)}(\bm r)\end{array}\right)\\
&=\left(\begin{array}{c}\tilde{\psi}_{\bm\gamma,b}^{(1)}(\bm r)\\ \tilde{\psi}_{\bm\gamma,t}^{(1)}(\bm r)\end{array}\right),
\label{C38}
\end{aligned}
\eeqn
which implies that $L_{\bm \gamma}^{(1)}$ is 0 in this case. $L_{\bm k}^{(n)}$ in other cases can be derived in a similar way (see Appendix~\ref{appendixII}).

\begin{table}[t]
\centering
\setlength\tabcolsep{1.5pt}
\renewcommand{\multirowsetup}{\centering}
\renewcommand{\arraystretch}{1.5}
\caption{ The angular momentum $L_{\bm k}^{(n)}$ of state $\psi_{\bm k}^{(n)}(\bm r)$ at high-symmetry momenta in phases (ii) and (iii). }
\begin{tabular}{|c|c|c|c|c|}
\hline
Phase&$(C_{+K}^{(1)},C_{+K}^{(2)})$&\diagbox[width=65pt,height=11mm]{$\bm k$}{$L_{\bm k}^{(n)}$} {states}& ${\psi}_{\bm k}^{(1)}(\bm r)$&${\psi}_{\bm k}^{(2)}(\bm r)$ \\
\hline
\multirow{3}{*}{\makecell{(ii)}}&\multirow{3}{*}{\makecell{(0,0)}}&$\bm \gamma$&$0$ & $1$ \\
\cline{3-5}
~&~&$ \bm \kappa$&$0$ & $1$ \\
\cline{3-5}
~&~&$ \bm \kappa^{\prime}$&$0$ & $1$ \\
\hline
\multirow{3}{*}{\makecell{(iii)}}&\multirow{3}{*}{\makecell{(1,-1)}}&$\bm \gamma$&$0$ & $1$ \\
\cline{3-5}
~&~&$ \bm \kappa$&$0$ & $1$ \\
\cline{3-5}
~&~&$ \bm \kappa^{\prime}$&$1$ & $0$ \\
\hline
\end{tabular}
\label{tab1}
\end{table}

In Table~\ref{tab1}, we list $L_{\bm k}^{(n)}$  of the first two moir\'e valence bands at the $C_3$ invariant momenta in phases (ii) and (iii). In phase (ii), $L_{\bm k}^{(n)}$ for a given $n \in \{1,2\}$ takes the same value, namely, $L_{\bm k}^{(1)}=0$ and $L_{\bm k}^{(2)}=1$ for $\bm k \in \{\bm \gamma, \bm \kappa, \bm \kappa'\}$.
In phase (iii), moir\'e bands have band inversion at $\bm\kappa^{\prime}$ point, which changes the values of $L_{\bm \kappa^{\prime}}^{(n)}$ to  $(L_{\bm \kappa^{\prime}}^{(1)},L_{\bm \kappa^{\prime}}^{(2)})=(1,0)$.  
The above analysis of $L_{\bm k}^{(n)}$ is consistent with the calculation of Chern number $C_{+K}^{(n)}$, since $[C_{+K}^{(n)}-(L_{\bm \gamma}^{(n)}+L_{\bm \kappa}^{(n)}+L_{\bm \kappa^{\prime}}^{(n)})]$ mod 3 $=0$ in a system with $C_3$ symmetry \cite{Fang2012}.

We now turn to the $M_x\mathcal{T}$ symmetry. For the Hamiltonian $\tilde{H}_{+}(\bm r)$, we note that $[\tilde{H}_{+}(-x,y)]^{*}=\tilde{H}_{+}(x,y)$. This identity implies that the
$M_x\mathcal{T}$ symmetry of Hamiltonian $H_{+}(\bm r)$ can be represented by $M_x\mathcal{T}=U^{-1}(\bm r)\mathcal{M}_x\mathcal{K}U(\bm r)$, where $\mathcal{M}_x$ is the operation that only flips $x$ to $-x$.

In the momentum space, the $M_x\mathcal{T}$ operator changes momentum $(k_x,k_y)$ to $(k_x,-k_y)$. Therefore, $E^{(n)}(k_x,k_y)=E^{(n)}(k_x,-k_y)$, where $E^{(n)}(\bm k)$ is the energy of state $\psi_{\bm k}^{(n)}$ under $H_{+}(\bm r)$. Moreover, the Bloch state $\psi_{\bm k}^{(n)}$ with $k_y=0$  is an eigenstate of  the $M_x\mathcal{T}$ symmetry, but the eigenvalue is gauge dependent since  $M_x\mathcal{T}$ is an antiunitary operator.

\begin{figure*}
\centering
\includegraphics[width=0.98\textwidth]{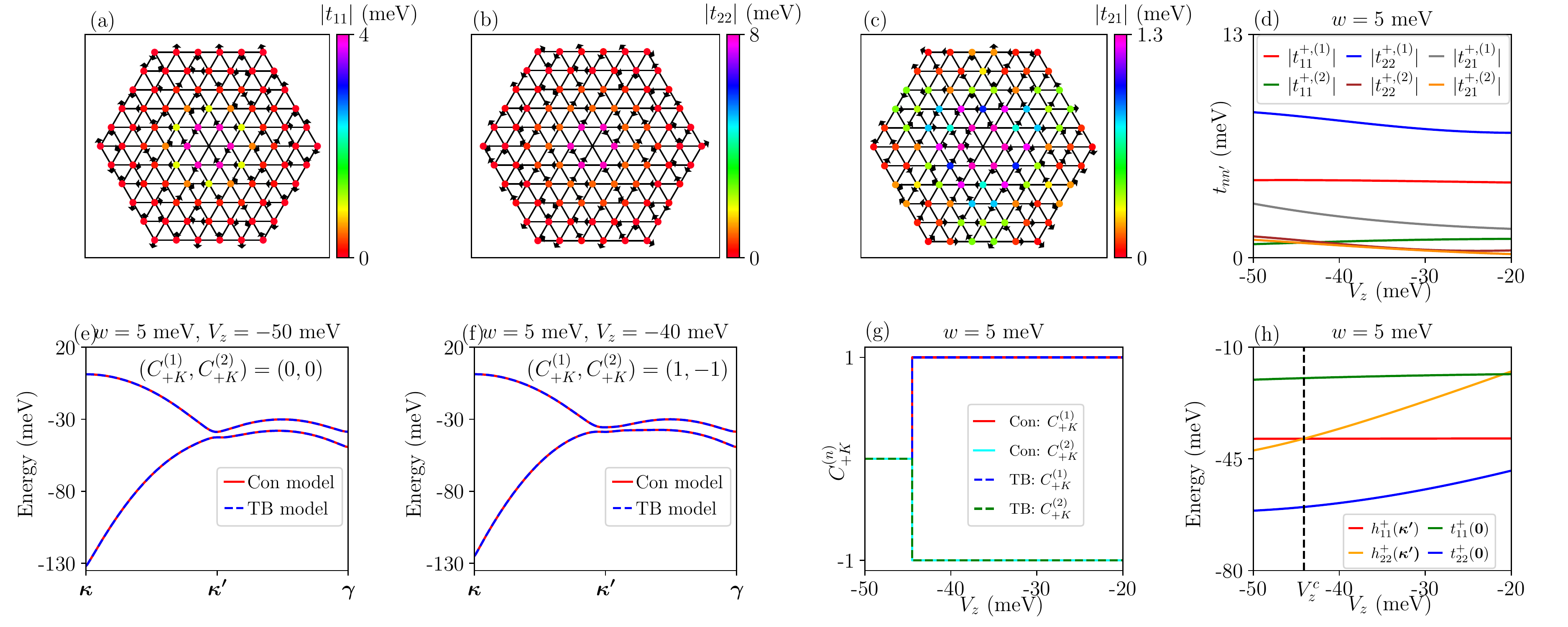}
\caption{ (a)-(c) The numerical values of hopping parameters $t_{11}^{+}(\bm R)$, $t_{22}^{+}(\bm R)$, and $t_{21}^{+}(\bm R)$ for $\bm R \neq \bm 0$. The color of the dots and the direction of the arrows at site $\bm R$ indicate the absolute value and the phase of $t_{nn^{\prime}}^{+}(\bm R)$, respectively.  We take the same model parameters as those used for Fig.~\ref{Chern}(f). (d) The absolute values of  nearest-neighbor ($|t_{nn^{\prime}}^{+,(1)}|$) and next nearest neighbor ($|t_{nn^{\prime}}^{+,(2)}|$) hopping parameters, as functions of $V_z$. (e), (f) Bands obtained from the TB model compared to those from the continuum model. (g) Chern numbers $C_{+K}^{(1)}$ and $C_{+K}^{(2)}$ given by the TB model compared to those given by the continuum model. (h) The numerical values of $h_{11}^{+}(\bm \kappa^{\prime})$, $h_{22}^{+}(\bm \kappa^{\prime})$, $t_{11}^{+}(\bm 0)$, and $t_{22}^{+}(\bm 0)$, as functions of $V_z$. The vertical black dashed line marks the topological phase transition point $V_z^{c}$. In (d), (g), and (h), we fix $w=5$ meV.}
\label{Values}
\end{figure*}

\section{Wannier states }
\label{IV}

We construct Wannier states for the first two moir\'e bands in phases (ii) and (iii), which is feasible because $C_{+ K}^{(1)}+C_{+ K}^{(2)}=0$ in both phases. The two phases are separated by a single topological phase transition with the band gap closing and reopening at $\bm \kappa'$ point.
Therefore, we can construct a unified TB model to describe the two phases.

The center of the Wannier states can be determined by $C_3$ eigenvalues at the high-symmetry momenta. We start with phase (ii), where the first and second bands are both topologically trivial, and therefore, can be separately described by a single-orbital TB model on a triangular lattice. In the first (second) band of phase (ii), the $C_3$ eigenvalues take the same value at $\bm \gamma,\bm \kappa$, and $\bm \kappa^{\prime}$ momenta, which implies that the Wannier center for the first (second) band is localized at MM sites (see Appendix~\ref{appendixIII}). By this argument, we can build a two-orbital TB model for the first two bands in phases (ii) and (iii), where the two Wannier orbitals are both localized at MM sites.

The Wannier states located at $\bm R= \bm 0$ (one of the MM sites) can be formally constructed as
\beqn
W^{(n)}(\bm r)&=&\frac{1}{\sqrt{N}}\sum_{\bm k}\phi_{\bm k}^{(n)}(\bm r),
\label{Wannier3}
\eeqn
where $n$ labels the two Wannier states $(n=1,2)$, $N$ is the number of moir\'e unit cells, and $\phi_{\bm k}^{(n)}(\bm r)$ is defined by
\beqn
\phi_{\bm k}^{(n)}(\bm r)=\sum_{n^{\prime}=1,2}V_{\bm k}^{n^{\prime} n}\psi_{\bm k}^{(n^{\prime)}}(\bm r).
\eeqn
Here the $2\times 2$ unitary matrix $V_{\bm k}$ is used to disentangle the layer hybridization. We determine $V_{\bm k}$ such that  ${\phi}_{\bm k}^{(1)}$ (${\phi}_{\bm k}^{(2)}$) is maximally polarized to the bottom (top) layer.
This maximum value problem can be transformed to seek the eigenbasis of the layer polarization operator $\sigma_z$ projected to the subspace spanned by $\psi_{\bm k}^{(1)}$ and $\psi_{\bm k}^{(2)}$,
\beqn
{\Pi}_{\bm k}=\begin{pmatrix}
\langle\psi_{\bm k}^{(1)}|\sigma_z|\psi_{\bm k}^{(1)}\rangle&\langle\psi_{\bm k}^{(1)}|\sigma_z|\psi_{\bm k}^{(2)}\rangle\\
\langle\psi_{\bm k}^{(2)}|\sigma_z|\psi_{\bm k}^{(1)}\rangle&\langle\psi_{\bm k}^{(2)}|\sigma_z|\psi_{\bm k}^{(2)}\rangle
\end{pmatrix}.
\eeqn
The desired $V_{\bm k}$ is given by
\beqn
&&V_{\bm k}^{\dagger}\Pi_{\bm k}V_{\bm k}=\begin{pmatrix}\rho_{\bm k}^{(1)}&0\\
0&\rho_{\bm k}^{(2)}\end{pmatrix},
\label{Wannier4}
\eeqn
where  $\rho_{\bm k}^{(1)}>\rho_{\bm k}^{(2)}$. We further fix the gauge such that the bottom (top) layer component of $\phi_{\bm k}^{(1)}$ ($\phi_{\bm k}^{(2)}$) is real and positive at $\bm r=0$.

The Wannier states constructed using the above procedures for the first two bands in Fig.~\ref{Chern}(f) are shown in Fig.~\ref{Wannier}, which plots both the amplitude and phase for each layer component of $W^{(n)}(\bm r) = [{W}_{b}^{(n)}(\bm r),{W}_{t}^{(n)}(\bm r)]^{T} $. The Wannier state $W^{(1)}(\bm r)$ mainly resides on the bottom layer, while $W^{(2)}(\bm r)$ has significant weights on both layers. 

The symmetry properties of the Wannier states can be analyzed in a similar way as that discussed in Sec.~\ref{III}. As illustrated in Fig.~\ref{Wannier}, the Wannier states are symmetric under $C_3$ symmetry with symmetry eigenvalues given by
\beqn
\begin{aligned}
&C_3W^{(1)}(\bm r)=W^{(1)}(\bm r),\\
&C_3W^{(2)}(\bm r)=e^{i2\pi/3}W^{(2)}(\bm r).
\label{Wannier6}
\end{aligned}
\eeqn
Thus,  $W^{(1)}(\bm r)$ and $W^{(2)}(\bm r)$ have angular momentum 0 and 1, respectively. 

By construction, the Wannier states have a gauge such that  $W^{(1)}_{b}(\bm r = \bm 0) >0 $ and $W^{(2)}_{t}(\bm r = \bm 0) >0 $. Under this gauge, both Wannier states are invariant under $M_x\mathcal{T}$ symmetry with symmetry eigenvalue 1,
\beqn
\begin{aligned}
& M_x\mathcal{T} W^{(1)}(\bm r)=W^{(1)}(\bm r),\\
& M_x\mathcal{T} W^{(2)}(\bm r)=W^{(2)}(\bm r).
\label{Wannier7}
\end{aligned}
\eeqn

Therefore, the constructed Wannier states are symmetric with respect to the $C_3$ and $M_x\mathcal{T}$ symmetries.
Finally, Wannier states located at a generic lattice site $\bm R$ are obtained through lattice translation, $W_{\bm R}^{(n)}(\bm r)=W^{(n)}(\bm r-\bm R)$.

\begin{figure}[b]
\centering
\includegraphics[width=0.4\textwidth]{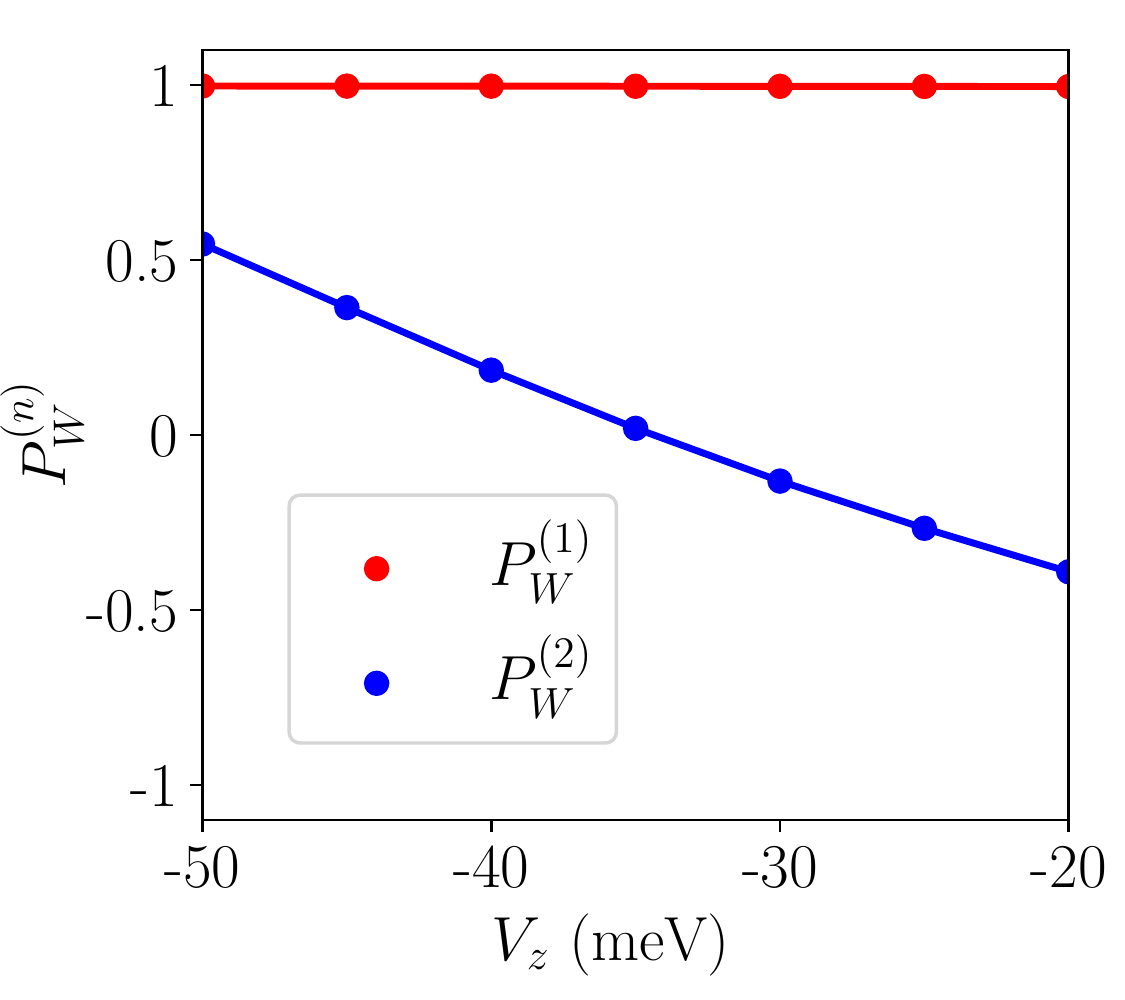}
\caption{The layer polarization of the two Wannier states as a function of $V_z$. The parameter values are the same as those used for Fig.~\ref{Chern}(f).} 
\label{WP}
\end{figure}

\section{Tight-binding model}
\label{V}

We further construct the TB model based on the obtained Wannier states,
\beqn
H_{\tau,\text{TB}}=\sum_{\bm R\bm R^{\prime}}\sum_{nn^{\prime}}t_{nn^{\prime}}^{\tau}(\bm R-\bm R^{'})c_{\bm R,\tau,n}^{\dagger}c_{\bm R^{\prime},\tau, n^{\prime}},
\label{TB1}
\eeqn
where $c_{\bm R,\tau,n}^{\dagger}$ ($c_{\bm R^{\prime},\tau,n^{\prime}}$) is the electron creation (annihilation) operator for the $n$th Wannier state in valley $\tau$ at the lattice position $\bm R$, and $t_{nn^{\prime}}^{\tau}$ is the hopping parameter. In Eq.~\eqref{TB1}, we reintroduce the valley index $\tau$ for completeness. Again, we first focus on the TB model in $+K$ valley. The hopping parameter is calculated in the following way
\beqn
t_{nn^{\prime}}^{+}(\bm R)&=&\langle W_{\bm R}^{(n)}|H_{+}|W_{\bm 0}^{(n^{\prime})}\rangle\nonumber\\
&=&\frac{1}{N}\sum_{\bm k}e^{i\bm k\cdot \bm R}\sum_{n^{\prime\prime}}[V_{\bm k}^{n^{\prime\prime} n}]^{*}E^{(n^{\prime\prime})}(\bm k)V_{\bm k}^{n^{\prime\prime} n^{\prime}},\nonumber\\
\label{TB2}
\eeqn
where $E^{(n^{\prime\prime})}(\bm k)$ is the energy of state $\psi_{\bm k}^{(n^{\prime\prime})}(\bm r)$ under $H_{+}(\bm r)$. 

The symmetries of the Hamiltonian $H_{+}$ and the Wannier states impose restrictions on the hopping parameters.
The hermiticity of the Hamiltonian requires that
\beqn
t_{nn^{\prime}}^{+}(\bm R)=[t_{n^{\prime}n}^{+}(-\bm R)]^{*}.
\label{TB4}
\eeqn
The $C_3$ symmetry leads to the following constraints,
\beqn
\begin{aligned}
&t_{11}^{+}(\bm R)=t_{11}^{+}(\hat{R}_{3}\bm R),\quad t_{22}^{+}(\bm R)=t_{22}^{+}(\hat{R}_{3}\bm R) ,\\
&t_{21}^{+}(\bm R)=e^{i2\pi/3}t_{21}^{+}(\hat{R}_{3}\bm R).
\label{TB3}
\end{aligned}
\eeqn
Finally, the $M_x\mathcal{T}$ symmetry imposes that
\beqn
t_{nn^{\prime}}^{+}(x,y)=[t_{nn^{\prime}}^{+}(-x,y)]^{*}.
\label{TB5}
\eeqn
At $\bm R=\bm 0$, Eqs.~\eqref{TB4} and \eqref{TB3} require that $t_{nn}^{+}(\bm 0)$ is real and $t_{12}^{+}(\bm 0)=t_{21}^{+}(\bm 0)=0$. Along $x=0$,  Eq.~\eqref{TB5}  requires that $t_{nn^{\prime}}^{+}(0,y)$ is real.

Figures \ref{Values}(a)-\ref{Values}(c) present the numerical values of the hopping parameters. It can be verified that the calculated $t_{nn^{\prime}}^{+}(\bm R)$ obey the aforementioned symmetry constraints in Eqs.~\eqref{TB4}, \eqref{TB3}, and \eqref{TB5}. In Fig.~\ref{Values}(d), 
we present the absolute values of nearest-neighbor $(|t_{nn^{\prime}}^{+,(1)}|)$ and next-nearest-neighbor $(|t_{nn^{\prime}}^{+,(2)}|)$ hopping parameters  as a function of $V_z$ at a fixed $w$; the numerical results show that  $|t_{11}^{+,(1)}|$ and $|t_{11}^{+,(2)}|$ remain almost constants with varying $V_z$, but other hopping parameters in Fig.~\ref{Values}(d) slowly decrease with the decreasing of $|V_z|$. The dependence of the hopping parameters on $V_z$, can be revealed by the layer polarization of the Wannier states, which is defined as
\beqn
P_W^{(n)}&=&\left\langle W^{(n)}\left|\sigma_z\right| W^{(n)}\right\rangle.
\eeqn
 As shown in Fig.~\ref{WP}, $P_W^{(1)}$ for the first Wannier state almost does not change with $V_z$ and is saturated to be $\sim 1$, indicating that the first Wannier state is primarily in the bottom layer. This explains the weak dependence of $t_{11}^{+,(1)}$ and $t_{11}^{+(2)}$ on $V_z$. In contrast, $P_W^{(2)}$ decreases with decreasing of $\left|V_z\right|$, which implies that the top layer component of $W^{(2)}$ becomes larger. The dependence of $t_{22}^{+,(1)}$ on $V_z$ is consistent with the variation of $W^{(2)}$ as a function of $V_z$.


\begin{figure*}
\centering
\includegraphics[width=0.98\textwidth]{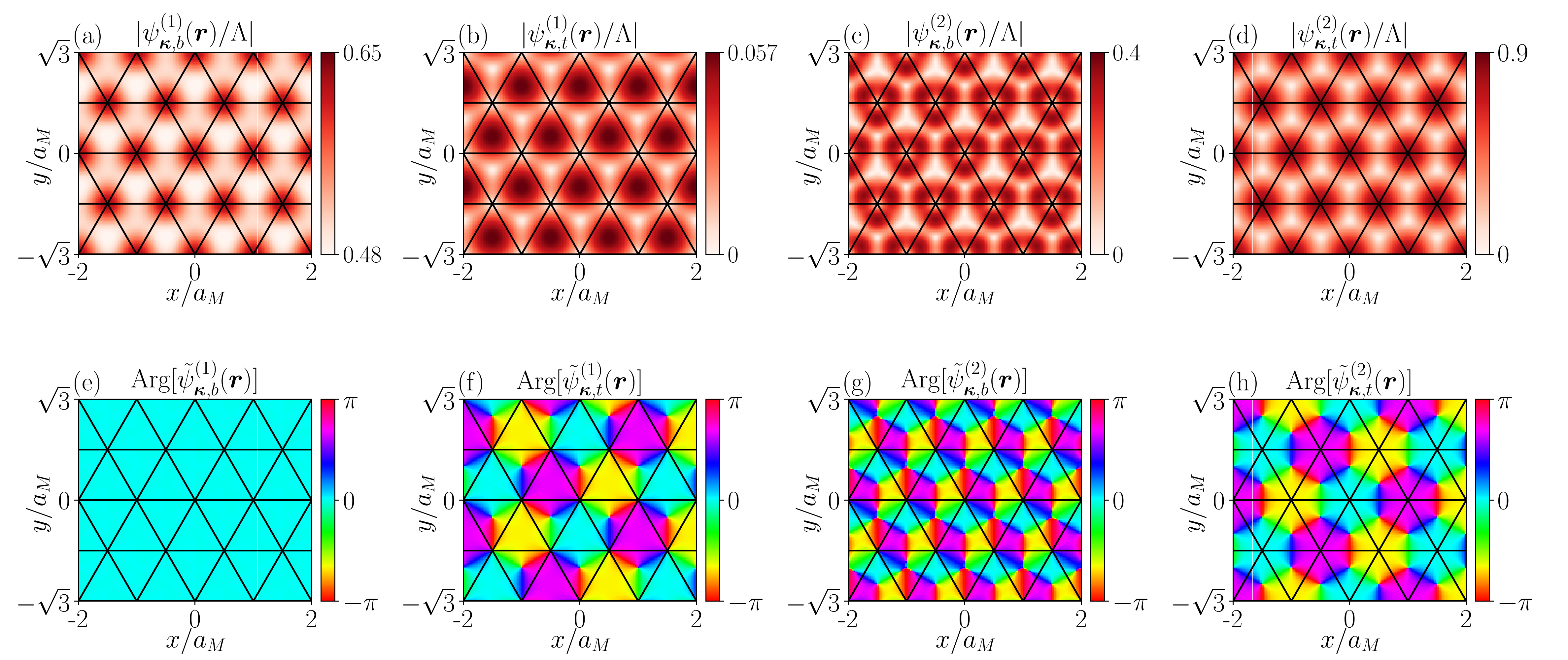}
\caption{The amplitude and phase of Bloch states ${\psi}_{\bm \kappa}^{(n)}(\bm r)=[{\psi}_{\bm \kappa,b}^{(n)}(\bm r),{\psi}_{\bm \kappa,t}^{(n)}(\bm r)]^{T}$ at $\bm \kappa$ point. Parameter values are the same as those used for Fig.~\ref{Chern}(f).}
\label{Bloch1}
\end{figure*}

The Bloch Hamiltonian obtained by performing  Fourier transformation to Hamiltonian $H_{+,\text{TB}}$ is given by
\beqn
H_{+,\text{TB}}(\bm k)=\begin{pmatrix} h_{11}^{+}(\bm k) &h_{12}^{+}(\bm k)\\
h_{21}^{+}(\bm k)&h_{22}^{+}(\bm k)\end{pmatrix}.
\label{TB6}
\eeqn
The matrix element $h_{nn^{\prime}}^{+}(\bm k)$ of Hamiltonian $H_{+,\text{TB}}(\bm k)$ can be written as
\beqn
\begin{aligned}
&h_{11}^{+}(\bm k)=t_{11}^{+}(\bm 0)+\sum_{\bm R\neq 0}t_{11}^{+}(\bm R)e^{-i\bm k\cdot\bm R},\\
&h_{22}^{+}(\bm k)=t_{22}^{+}(\bm 0)+\sum_{\bm R\neq 0}t_{22}^{+}(\bm R)e^{-i\bm k\cdot\bm R},\\
&h_{21}^{+}(\bm k)=\sum_{\bm R\neq 0}t_{21}^{+}(\bm R)e^{-i\bm k\cdot\bm R},
\label{TB7}
\end{aligned}
\eeqn
and $h_{12}^{+}(\bm k)=[h_{21}^{+}(\bm k)]^{*}$ owing to the hermiticity of Hamiltonian. By combining Eqs.~\eqref{TB2} and \eqref{TB7}, we can simplify $H_{+,\text{TB}}(\bm k)$ to be
\beqn
H_{+,\text{TB}}(\bm k)=V_{\bm k}^{\dagger}\begin{pmatrix}  E^{(1)}(\bm k) & 0 \\
0 & E^{(2)}(\bm k) \end{pmatrix} V_{\bm k}.
\label{HTBV}
\eeqn

Figures \ref{Values}(e) and \ref{Values}(f) plot the energy bands obtained from $H_{+,\text{TB}}(\bm k)$, which accurately reproduce the moir\'e bands in Fig.~\ref{Chern}(e) and Fig.~\ref{Chern}(f), respectively. 
The Chern numbers calculated using the TB model in Eq.~\eqref{HTBV} and the continuum model in Eq.~\eqref{Ham} are compared in Fig.~\ref{Values}(g), which confirms that the constructed TB model can faithfully describe the topological phase transition tuned by $V_z$.

The topological phase transition of $H_{+,\text{TB}}(\bm k)$ can also be understood by the $V_z$-tuned band inversion at $\bm \kappa'$ point. The band gap  at $\bm \kappa'$ closes when $h_{11}^{+}(\bm \kappa^{\prime})=h_{22}^{+}(\bm \kappa^{\prime})$ because the off-diagonal term $h_{12}^{+}(\bm \kappa^{\prime})$ vanishes. The diagonal terms $h_{nn}^{+}(\bm \kappa^{\prime})$ as functions of $V_z$ are presented in Fig.~\ref{Values}(h), which verifies the band gap closing at the topological phase transition. Figure~\ref{Values}(h) shows that $h_{11}^{+}(\bm \kappa^{\prime})$ is almost a constant as a function of $V_z$, but $V_z$ significantly tunes the onsite potential $t_{22}^{+}(\bm 0)$ of the second Wannier state $W^{(2)}(\bm r)$ and therefore,  $h_{22}^{+}(\bm \kappa^{\prime})$. This is because $V_z$ only tunes the top layer potential in Eq.~\eqref{Ham}.

Finally, we discuss the Wannier states and the TB model in the other valley. In Appendix~\ref{Appendix IV}, we explicitly construct the two Wannier states in $-K$ valley using the same procedure and gauge choice discussed in Sec.~\ref{IV},  and show that they can be expressed as $-\mathcal{T}W^{(1)}(\bm r)$ and $\mathcal{T}W^{(2)}(\bm r)$, respectively. Therefore, the $\mathcal{T}$ symmetry relates the hopping parameters of the TB models in the two valleys in the following way,
\beqn
\begin{aligned}
&t_{11}^{-}(\bm R)=[t_{11}^{+}(\bm R)]^{*},\quad t_{22}^{-}(\bm R)=[t_{22}^{+}(\bm R)]^{*},\\
& t_{12}^{-}(\bm R)=-[t_{12}^{+}(\bm R)]^{*},
\end{aligned}
\eeqn
which fully determines the TB model in $-K$ valley.

\section{Discussion and Conclusion}
\label{VI}
In summary, symmetry-adapted Wannier states and TB model are constructed for the quantum spin Hall bands in AB-stacked MoTe$_2$/WSe$_2$. For each valley, the TB model is defined on a triangular lattice with two Wannier states on each lattice site. The two Wannier states have the same Wannier center but different angular momenta. The difference in the angular momenta of the two Wannier states is crucial for the topological phase transition induced by the displacement field. The constructed TB model is similar to the Bernevig-Hughes-Zhang model with band inversion between $s$-type and $p$-type orbitals \cite{Bernevig2006}. We emphasize that symmetry representation of the Bloch states at high-symmetry momenta essentially determines the Wannier centers.

Previously, the TB model for topological bands in twisted TMD bilayers has been shown to be a generalized Kane-Mele model \cite{Kane2005,Kane2005a} on a honeycomb lattice for certain model parameters \cite{Wu2019, Pan2020, Devakul2021, Rademaker2022Spin}. Our study shows that the TB model for topological bands depends on system details, and should be constructed case by case based on symmetry analysis of Bloch states. 
The developed methods to analyze the symmetry of moir\'e
Hamiltonian and construct Wannier states are applicable to other TMD moir\'e systems.

\begin{figure*}
\centering
\includegraphics[width=0.98\textwidth]{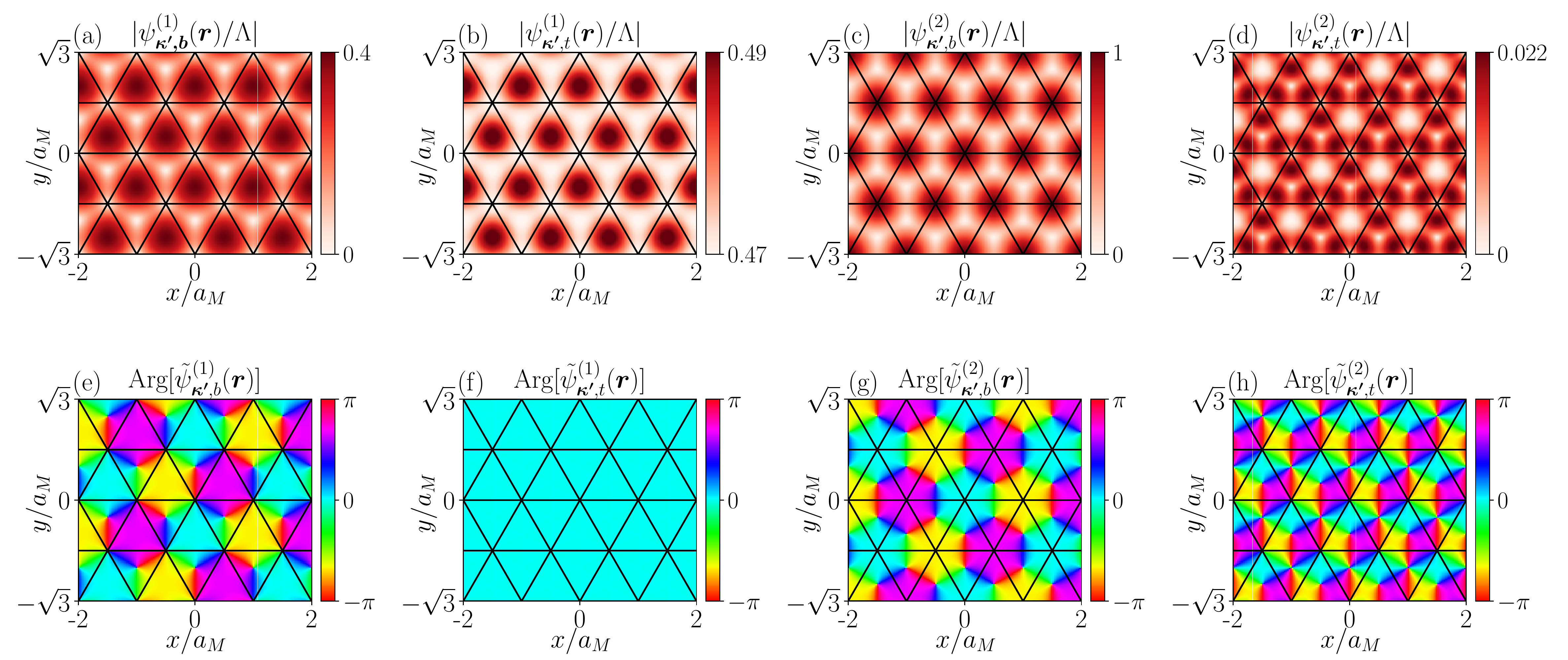}
\caption{The amplitude and phase of Bloch states ${\psi}_{\bm \kappa^{\prime}}^{(n)}(\bm r)=[{\psi}_{\bm \kappa^{\prime},b}^{(n)}(\bm r),{\psi}_{\bm \kappa^{\prime},t}^{(n)}(\bm r)]^{T}$ at $\bm \kappa^{\prime}$ point. Parameter values are the same as those used for Fig.~\ref{Chern}(f).}
\label{Bloch2}
\end{figure*}

We also construct the maximally localized Wannier states (see Appendix~\ref{Appendix V} for details), which have less spread in real space but are qualitatively similar to the Wannier states before optimization. We expect that the constructed Wanner states and TB model can provide a basis to study the rich interaction-driven quantum phase diagrams in AB-stacked MoTe$_2$/WSe$_2$.

\section{ACKNOWLEDGMENTS}
F. W. thanks H. Pan and R.-X. Zhang for helpful discussions. This work is supported by National Natural Science Foundation of China (Grant No. 12274333), National Key Research and Development Program of China (Grant No. 2021YFA1401300), and start-up funding of Wuhan University.

\begin{appendix}

\section{Time-reversal symmetry }
\label{appendixI}
The moir\'e Hamiltonian of AB-stacked MoTe$_{2}$/WSe$_{2}$ can be expressed in the second quantized form as follows,
\beqn
\hat{\mathcal{H}}_0=\int d^2\bm r\Phi^{\dagger}(\bm r)H_0(\bm r)\Phi(\bm r),
\label{A0}
\eeqn
where
\beqn
&&{H}_0(\bm r)=\begin{pmatrix}
H_{+}(\bm r)&0\\
0&H_{-}(\bm r)\end{pmatrix},\nonumber\\
&&H_{\tau}(\bm r)=\begin{pmatrix}
\mathcal{H}_{b,\tau}(\bm r)+\Delta_{{b}}(\bm{r}) & \Delta_{{T},\tau}(\bm{r})\\
\Delta_{{T},\tau}^\dag(\bm{r}) & \mathcal{H}_{t,\tau}(\bm r)+ \Delta_{t}(\bm{r})+V_{z}
\end{pmatrix}, \nonumber\\
&&{\Phi}(\bm r)=(\varphi_{+,b,\uparrow}(\bm r) ,\,\,\,\varphi_{+,t,\downarrow}(\bm r),\,\,\,\varphi_{-,b,\downarrow}(\bm r), \,\,\,\varphi_{-,t,\uparrow}(\bm r))^{T}.\nonumber\\
\label{A1}
\eeqn
In Eq.~\eqref{A1}, $\varphi_{\tau,l,s}(\bm r )$ ( $\varphi_{\tau,l,s}^{\dagger}(\bm r )$) is the electron annihilation (creation) operator, where $\tau=\pm$ is the valley index,   $l=b,t$ is the layer index, and $s=\uparrow, \downarrow$ is the spin index. The time-reversal symmetry acts on  $\varphi_{\tau,l,s}(\bm r )$ in the following way
\beqn
\begin{aligned}
&\mathcal{T}\varphi_{+,b,\uparrow}(\bm r)\mathcal{T}^{-1}=-\varphi_{-,b,\downarrow}(\bm r),\\
&\mathcal{T}\varphi_{+,t,\downarrow}(\bm r)\mathcal{T}^{-1}=\varphi_{-,t,\uparrow}(\bm r),\\
&\mathcal{T}\varphi_{-,b,\downarrow}(\bm r)\mathcal{T}^{-1}=\varphi_{+,b,\uparrow}(\bm r),\\
&\mathcal{T}\varphi_{-,t,\uparrow}(\bm r)\mathcal{T}^{-1}=-\varphi_{+,t,\downarrow}(\bm r).
\label{A2}
\end{aligned}
\eeqn
Therefore, the time-reversal symmetry can be written as $\mathcal{T}=i\tau_y\sigma_z\mathcal{K}$ in the basis ${\Phi}(\bm r)$ and acts on the Hamiltonian $H_0$ as
\beqn
\mathcal{T}H_0(\bm r)\mathcal{T}^{-1}&=&\begin{pmatrix}
\sigma_z\mathcal{K}H_{-}(\bm r)\sigma_z\mathcal{K}&0\\
0&\sigma_z\mathcal{K}H_{+}(\bm r)\sigma_z\mathcal{K}\end{pmatrix}\nonumber\\
&=&H_0(\bm r).
\label{A3}
\eeqn

\section{Angular momentum of Bloch states }
\label{appendixII}

In Fig.~\ref{Bloch1}, we present the amplitude and phase for each layer component of ${\psi}_{\bm \kappa}^{(n)}(\bm r)$, which represents the Bloch states of the first two bands $(n=1,2)$ in Fig.~\ref{Chern}(f) at $\bm \kappa$ point. Figure~\ref{Bloch2} is similar to Fig.~\ref{Bloch1}, but for the Bloch states ${\psi}_{\bm \kappa'}^{(n)}(\bm r)$ at $\bm \kappa'$ point. The angular momentum under $C_3$ symmetry for Bloch states shown in Figs.~\ref{Bloch1} and \ref{Bloch2} can be analyzed using the approach discussed in Sec.~\ref{III}, with results given by,
\beqn
\begin{aligned}
&{C}_3{\psi}_{\bm \kappa}^{(1)}(\bm r)={\psi}_{\bm \kappa}^{(1)}(\bm r),\\
&{C}_3{\psi}_{\bm \kappa}^{(2)}(\bm r)=e^{i2\pi/3}{\psi}_{\bm \kappa}^{(2)}(\bm r),\\
&{C}_3{\psi}_{\bm \kappa^{\prime}}^{(1)}(\bm r)=e^{i2\pi/3}{\psi}_{\bm\kappa^{\prime}}^{(1)}(\bm r),\\
&{C}_3{\psi}_{\bm \kappa^{\prime}}^{(2)}(\bm r)={\psi}_{\bm \kappa^{\prime}}^{(2)}(\bm r),
\label{A7}
\end{aligned}
\eeqn
which gives rise to the angular momentum listed in Table~\ref{tab1}.

\section{Wannier center}
\label{appendixIII}

For AB-stacked MoTe$_2$/MoSe$_2$, three are three high-symmetry positions in the moir\'e superlattices, namely MM, XX, and MX points. A Wannier state can be centered at one of these three points. For different choices of the Wannier center, the corresponding Bloch state at high-symmetry momenta $\bm \gamma$, $\bm \kappa$, and $\bm \kappa^{\prime}$ has different patterns of the $C_3$ symmetry eigenvalues. 

\begin{table}[b]
\centering
\setlength\tabcolsep{4pt}
\renewcommand{\multirowsetup}{\centering}
\renewcommand{\arraystretch}{2}
\caption{ The angular momentum $L_{\bm k}$ at the high-symmetry momenta for Wannier states centered at different positions.}
\begin{tabular}{|c|c|c|c|}
\hline
\diagbox[width=60pt,height=12mm]{$\bm k$}{$L_{\bm k}$} {$\bm{r}_{\alpha}$}& \makecell{MM\\ $\bm r_1=(0,0)$}&$\makecell{\text{XX}\\ \bm r_{2}=\\ a_M(1/2,1/2\sqrt{3})}$&$ \makecell{\text{MX}\\  \bm r_{3}=\\ a_M(1,1/\sqrt{3}) }$\\
\hline
$\bm \gamma$&$\ell$&$\ell$&$\ell$ \\
\cline{1-4}
$\bm \kappa$&$\ell$&$\ell+1$&$\ell-1$ \\
\cline{1-4}
$\bm \kappa^{\prime}$&$\ell$&$\ell-1$&$\ell+1$ \\
\hline
\end{tabular}
\label{tab3}
\end{table}

We consider a Wannier state $\chi(\bm r-\bm R-\bm r_{\alpha})$ centered at $\bm R+\bm r_{\alpha}$, where $\bm R$ is the lattice translation vector and $\bm r_{\alpha}$ represents one of the three positions, namely, $\bm r_{\text{1}}=(0,0)$ for the MM site, $\bm r_{2}=a_M(1/2,1/2\sqrt{3})$ for the XX site
, and $\bm r_{3}=a_M(1,1/\sqrt{3})$ for the MX site. The corresponding Bloch state can be written as
\beqn
\Psi_{\bm k}(\bm r)=\sum_{\bm R}e^{i\bm k\bm \cdot (\bm R+\bm r_{\alpha})}\chi(\bm r-\bm R-\bm r_{\alpha}).
\eeqn
 
The threefold rotation symmetry $\mathcal{C}_3$ acts on the Bloch state $\Psi_{\bm k}(\bm r)$ as
\beqn
\mathcal{C}_3\Psi_{\bm k}(\bm r)&=&\sum_{\bm R}e^{i\bm k\bm \cdot (\bm R+\bm r_{\alpha})}\mathcal{D}_{\mathcal{C}_3}\chi(\hat{R}_3\bm r-\bm R-\bm r_{\alpha})\nonumber\\
&=&\sum_{\bm R}e^{i\bm k\bm \cdot (\bm R+\bm r_{\alpha})}\mathcal{D}_{C_3}\chi(\hat{R}_3[\bm r-\hat{R}_3^{-1}(\bm R+\bm r_{\alpha})])\nonumber\\
&=&\sum_{\bm R}e^{i\bm k\bm \cdot(\bm R+\bm r_{\alpha})}\mathcal{D}_{C_3}\chi(\hat{R}_3[\bm r-\bm R^{\prime}-\bm r_{\alpha}])\nonumber\\
&=&e^{i2\pi\ell /3}\sum_{\bm R}e^{i\bm k\bm \cdot (\bm R+\bm r_{\alpha})}\chi(\bm r-\bm R^{\prime}-\bm r_{\alpha}),\nonumber\\
\label{A10}
\eeqn
where  $\bm R^{\prime}+\bm r_{\alpha}=\hat{R}_3^{-1}(\bm R+\bm r_{\alpha})$, $\mathcal{D}_{\mathcal{C}_3}$ is the representation matrix of $\mathcal{C}_3$ operation, and $\ell$ is the angular momentum of Wannier state $\chi$.  Equation~\eqref{A10} can be further written as
\beqn
\mathcal{C}_3\Psi_{\bm k}(\bm r)&=&e^{i2\pi\ell /3}\sum_{\bm R^{\prime}}e^{i\bm k\bm \cdot \hat{R}_3(\bm R^{\prime}+\bm r_{\alpha})}\chi(\bm r-\bm R^{\prime}-\bm r_{\alpha})\nonumber\\
&=&e^{i2\pi\ell /3}\sum_{\bm R^{\prime}}e^{i(\hat{R}_3^{-1}\bm k)\bm \cdot (\bm R^{\prime}+\bm r_{\alpha})}\chi(\bm r-\bm R^{\prime}-\bm r_{\alpha}).\nonumber\\
\label{A11}
\eeqn

\begin{figure*}
\centering
\includegraphics[width=0.98\textwidth]{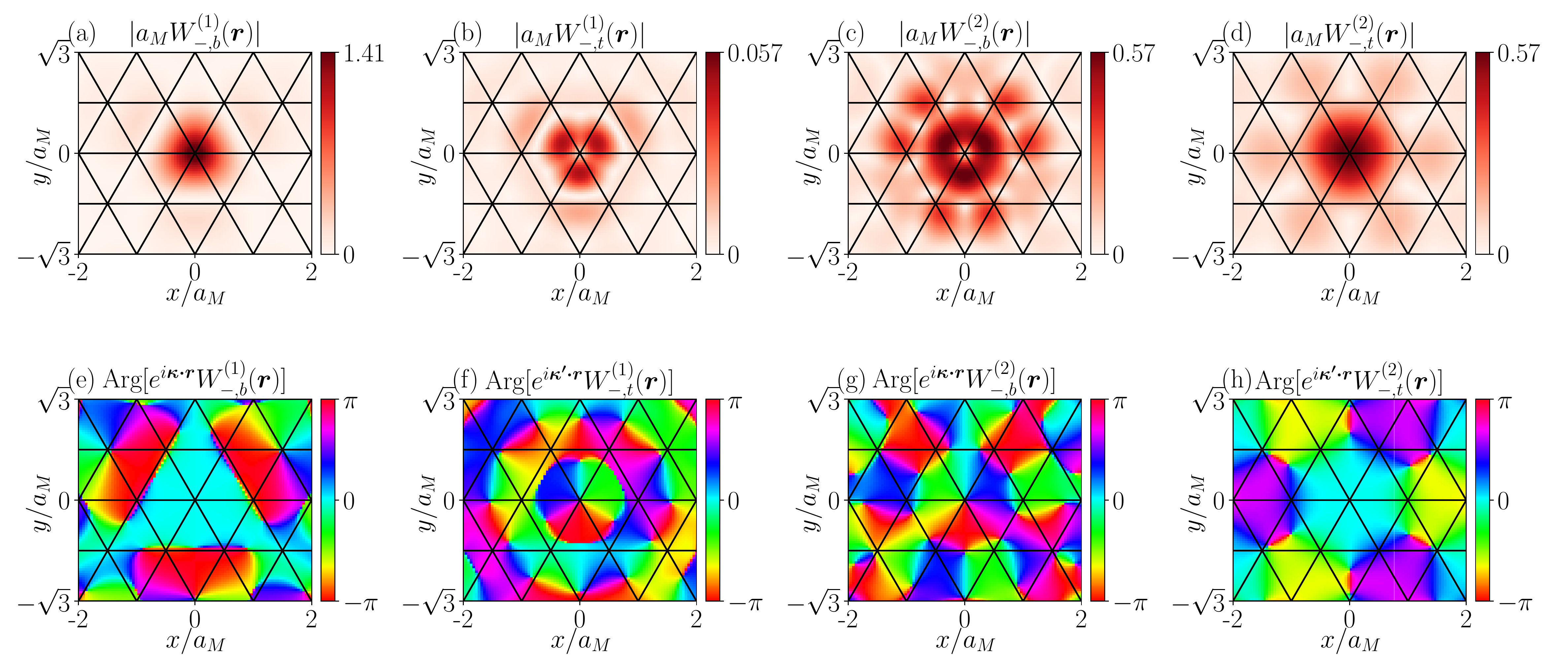}
\caption{The amplitude and phase of Wannier states ${W}_{-}^{(n)}(\bm r)=[{W}_{-,b}^{(n)}(\bm r),{W}_{-,t}^{(n)}(\bm r)]^{T}$ in $-K$ valley. (a)-(d) The amplitude of ${W}_{-,l}^{(n)}(\bm r)$. (e)-(h) The phase of $e^{i\bm \kappa\cdot\bm r}{W}_{-,b}^{(n)}(\bm r)$ and  $e^{i\bm \kappa^{\prime}\cdot\bm r}{W}_{-,t}^{(n)}(\bm r)$. We take the gauge such that $W_{-,b}^{(1)}(\bm r)$ and $W_{-,t}^{(2)}(\bm r)$ are real and positive at $\bm r=0$. Parameter values are the same as those used for Fig.~\ref{Chern}(f).}
\label{Wannier1}
\end{figure*}

At the high-symmetry momenta $\bm k=\bm\gamma,\bm\kappa,\bm\kappa^{\prime}$, we have $\hat{R}_3^{-1}\bm k=\bm k+{\bm g}_{\bm k}$, where ${\bm g}_{\bm k}$ is a reciprocal lattice vector,  with ${\bm g}_{\bm \gamma}=\bm 0$ for $\bm k=\bm \gamma$, ${\bm g}_{\bm \kappa}=\frac{4\pi}{\sqrt{3}a_M}(\sqrt{3}/2,-1/2)$ for $\bm k=\bm \kappa$, and ${\bm g}_{\bm \kappa'}=\frac{4\pi}{\sqrt{3}a_M}(0,-1)$ for $\bm k=\bm \kappa^{\prime}$. 
At these three high-symmetry points,  Eq.~\eqref{A11} can be further written as
\beqn
\mathcal{C}_3\Psi_{\bm k}(\bm r)&=&e^{i2\pi\ell /3}\sum_{\bm R^{\prime}}e^{i(\bm k+{\bm g}_{\bm k})\bm \cdot (\bm R^{\prime}+\bm r_{\alpha})}\chi(\bm r-\bm R^{\prime}-\bm r_{\alpha})\nonumber\\
&=&e^{i2\pi\ell /3}e^{i\bm {g_k \cdot \bm r_{\alpha}}}\sum_{\bm R^{\prime}}e^{i\bm k\bm \cdot (\bm R^{\prime}+\bm r_{\alpha})}\chi(\bm r-\bm R^{\prime}-\bm r_{\alpha})\nonumber\\
&=&e^{i2\pi\ell /3}e^{i\bm {g_k \cdot \bm r_{\alpha}}}\Psi_{\bm k}(\bm r)\nonumber\\
&=&e^{i2\pi L_{\bm k}/3}\Psi_{\bm k}(\bm r),
\label{A12}
\eeqn
where $L_{\bm k}=(\ell+3\bm {g_k \cdot r_{\alpha}}/2\pi)$ mod 3 is the angular momentum of the Bloch state $\Psi_{\bm k}$ under threefold rotation. In Table~\ref{tab3}, we list the angular momentum $L_{\bm k}$ at the high-symmetry momenta for different positions of the Wannier center.

As shown in Table~\ref{tab3}, $L_{\bm k}$ takes the same value at $\bm \gamma, \bm \kappa$, and $\bm \kappa'$ points for Wannier center at the MM site, but different values for Wannier center at the XX (MX) site. Therefore, only Wannier states centered at the MM sites are compatible with the pattern of angular momentum listed in Table~\ref{tab1}.

\section{Wannier states of $-K$ valley}
\label{Appendix IV}
In this section, we present the Wannier states in $-K$ valley and show how time-reversal symmetry relates the Wannier states from $\pm K$ valleys.

For definiteness, we use $W_{+}^{(n)}(\bm r)$ and $W_{-}^{(n)}(\bm r)$ to denote the Wannier states at $+K$ and $-K$ valleys, respectively. The Wannier state $W_{\tau}^{(n)}(\bm r)$ can be represented by a two-component spinor $[W_{\tau,b}^{(n)}(\bm r), W_{\tau,t}^{(n)}(\bm r)]^{T}$ in the layer pseudospin space. 
If the valley degree of freedom is also taken into account,  $W_{\tau}^{(n)}(\bm r)$ is then represented by a four-component spinor in the combined layer and valley pseudospin space,
\beqn
\begin{aligned}
W_{+}^{(n)}(\bm r) & = [W_{+,b}^{(n)}(\bm r), W_{+,t}^{(n)}(\bm r),0,0]^{T},\\
W_{-}^{(n)}(\bm r) & = [0,0,W_{-,b}^{(n)}(\bm r), W_{-,t}^{(n)}(\bm r)]^{T},
\end{aligned}
\eeqn
where we take the same basis as that for Hamiltonian ${H}_0$ in Eq.~\eqref{A1}.

We calculate the Wannier states $W_{-}^{(n)}(\bm r)$ in $-K$ valley using the same approach as presented in  Sec.~\ref{IV}. We also use the same gauge such that $W_{-,b}^{(1)}(\bm r)$ and $W_{-,t}^{(2)}(\bm r)$ are real and positive at $\bm r=0$. Figure~\ref{Wannier1} shows the calculated results for $W_{-}^{(n)}(\bm r)$.
It can be shown that $W_{-}^{(n)}(\bm r)$ also satisfies the $C_3$ and $M_x\mathcal{T}$ symmetries, and the angular momentum of $W_{-}^{(n)}(\bm r)$ is $0$ and $-1$ for $n=1$ and $2$, respectively.

We now turn to the time-reversal symmetry $\mathcal{T}$, which acts on the Wannier states $W_{+}^{(n)}(\bm r)$ as
\beqn
\begin{aligned}
\mathcal{T}W_{+}^{(n)}(\bm r)&=\begin{pmatrix}0&0&\mathcal{K}&0\\
0&0&0&-\mathcal{K}\\
-\mathcal{K}&0&0&0\\
0&\mathcal{K}&0&0\end{pmatrix}\left(\begin{array}{c}W_{+,b}^{(n)}(\bm r)\\ W_{+,t}^{(n)}(\bm r)\\ 0\\ 0\end{array}\right)\\
&=\left[0,0,-\mathcal{K}W_{+,b}^{(n)}(\bm r), \mathcal{K}W_{+,t}^{(n)}(\bm r)\right]^{T}.
\end{aligned}
\label{D1}
\eeqn
By comparing Figs.~\ref{Wannier} and \ref{Wannier1}, it can be verified following Eq.~\eqref{D1} that
\beqn
\begin{aligned}
&\mathcal{T}W_{+}^{(1)}(\bm r)=-W_{-}^{(1)}(\bm r),\\
&\mathcal{T}W_{+}^{(2)}(\bm r)=W_{-}^{(2)}(\bm r),
\end{aligned}
\label{D3}
\eeqn
which confirms that Wannier states from $\pm K$ valleys are connected by the $\mathcal{T}$ symmetry.

\begin{figure*}
\centering
\includegraphics[width=0.98\textwidth]{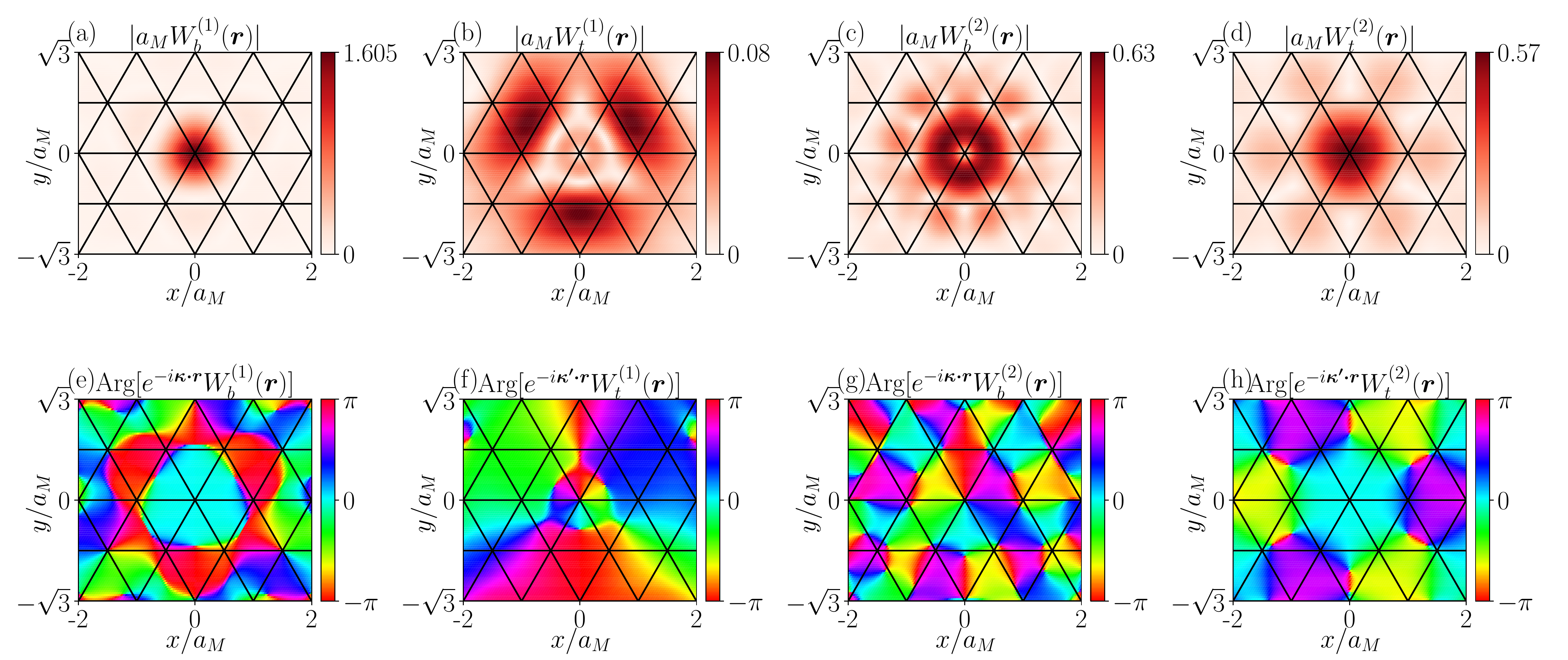}
\caption{The amplitude (a)-(d) and phase (e)-(h) of maximally localized Wannier states at $+K$ valley.  Parameter values are the same as those used for Fig.~\ref{Chern}(f).}
\label{maximally}
\end{figure*}

\section{Maximally localized Wannier states}
\label{Appendix V}
We construct the maximally localized Wannier states by following the method in Refs.~\onlinecite{Marzari1997,Marzari2012}. We take the Wannier states presented in Sec.~\ref{IV} as the initial guess, and then minimize the spread of the Wannier states. The obtained maximally localized Wannier states are illustrated in Fig.~\ref{maximally}. It can be verified that the maximally localized Wannier states remain symmetric under the $C_3$ and $M_x\mathcal{T}$ symmetries. 

The spread of the Wannier states is characterized by
\beqn
\Omega=\sum_{n=1,2}\langle W^{(n)}|\bm r^2|W^{(n)}\rangle-(\langle W^{(n)}|\bm r|W^{(n)}\rangle)^2,
\eeqn
where $W^{(n)}$ denotes the Wannier states. The spread $\Omega$ is $1.744 a_M^2$ before the optimization and reduced to $1.347 a_M^2$ after the optimization for parameters used in Fig.~\ref{Wannier}. As shown in Figs.~\ref{Wannier} and \ref{maximally}, the optimization only leads to quantitative changes in the Wannier states.

\end{appendix}

\bibliography{reference}

\end{document}